\documentclass[twocolumn,twocolappendix]{openj}

\usepackage{newtxtext,newtxmath}
\usepackage[T1]{fontenc}

\DeclareRobustCommand{\VAN}[3]{#2}
\let\VANthebibliography\thebibliography
\def\thebibliography{\DeclareRobustCommand{\VAN}[3]{##3}\VANthebibliography}

\usepackage{graphicx}	% Including figure files
\usepackage{amsmath}
\usepackage{ulem}% Advanced maths commands

\usepackage{orcidlink}
\usepackage{multirow}
\usepackage[nosort,capitalize]{cleveref}
\usepackage{xspace}
\usepackage{subcaption} % for subfigures
\usepackage{booktabs}

\newcommand{\borg}{\texttt{BORG}\xspace}

\newcommand{\slow}{\texttt{SLOW}\xspace}
\newcommand{\swift}{\texttt{SWIFT}\xspace}
\newcommand{\tmpp}{\texttt{2M++}\xspace}

\newcommand{\lcdm}{$\Lambda$CDM\xspace}
\newcommand{\msol}{M$_\odot$\xspace}

\newcommand{\manticore}{\texttt{Manticore}\xspace}
\newcommand{\manticorelocal}{\texttt{Manticore-Local}\xspace}

\Crefname{figure}{Figure}{Figures}
\Crefname{equation}{Equation}{Equations}
\Crefname{table}{Table}{Tables}
\Crefname{section}{Section}{Sections}

\begin{document}
\title{The Manticore-Local Cluster Catalogue: A Posterior Map of Massive Structures in the Nearby Universe}

\author{Stuart McAlpine\orcidlink{0000-0002-8286-7809}}
\email{stuart.mcalpine@fysik.su.se}
\affiliation{The Oskar Klein Centre, Department of Physics, Stockholm University, Albanova University Center, 106 91 Stockholm, Sweden}

\begin{abstract}
We present a publicly available catalogue of massive structures in the nearby Universe, constructed from the \textsc{Manticore-Local} posterior ensemble---a Bayesian field-level reconstruction that infers the underlying dark matter distribution from 2M++ galaxies. We identify massive structures by clustering the central haloes \textit{inferred} at $z = 0$ across the 80 posterior realizations, selecting at most one member per realization. These \textit{associations} serve as probabilistic counterparts to individual massive clusters, each with robust posterior estimates of mass, position, and velocity. The fiducial catalogue contains 401 associations with $\langle M_{200} \rangle \geq 10^{14}$~M$_{\odot}$, ambiguity rates below 5\%, and at least 20 member haloes across the posterior ensemble. We independently validate these systems through stacked \textit{Planck} thermal Sunyaev--Zel'dovich measurements, which yield significant detections ($>3\sigma$) showing the expected mass trend, consistent with the established $Y$--$M$ scaling relation. Many associations exhibit coherent evolutionary histories, meaning that their progenitor haloes across the posterior ensemble trace consistent merger pathways rather than diverging into unrelated assembly scenarios. Even with only $z = 0$ constraints, the inference narrows each system's possible assembly pathway: progenitor haloes at earlier times occupy Lagrangian volumes 2--5 times smaller than those of mass-matched haloes in unconstrained $\Lambda$CDM simulations, reflecting genuine information gain from the observational constraints. Cross-matches with X-ray catalogues reveal systematic mass-scale differences that align with known observational biases: \textsc{Manticore-Local} masses are typically 1.7 times ROSAT-based estimates but agree at unity with weak-lensing-calibrated eROSITA measurements. This illustrates the catalogue's potential to flag individual systems with significant mass discrepancies for targeted investigation. The resulting catalogue provides an observationally consistent map of massive structures in the local Universe, enabling direct cross-probe comparisons, hybrid analyses combining simulated and observed quantities, and systematic mass-scale studies. All data products are publicly available via \href{https://cosmictwin.org}{\texttt{cosmictwin.org}} as a community resource.
\end{abstract}


\maketitle

%%%%%%%%%%%%%%%%%%%%%%%%%%%%%%%%%%%%%%%%%%%%%%%%%%

%%%%%%%%%%%%%%%%% BODY OF PAPER %%%%%%%%%%%%%%%%%%

\section{INTRODUCTION}

Galaxy clusters are the most massive gravitationally bound structures in the Universe, forming at the intersections of cosmic filaments where matter flows converge \citep{Press1974}. As dominant nodes of the cosmic web, clusters probe both cosmology and astrophysics: their abundance and clustering trace the growth of structure \citep[e.g.][]{2011ARA&A..49..409A}, while their deep, dark-matter--dominated potential wells and dense galaxy populations make them key laboratories for galaxy evolution \citep{2012ARA&A..50..353K}. In the local Universe (within a few hundred Mpc, $z \lesssim 0.05$), iconic systems such as Virgo \citep{Ferrarese2012,Boselli2014} and Coma \citep{Rines2003,Kubo2007} can be studied in exceptional detail, providing stringent tests of dark matter, galaxy formation, and large-scale dynamics within the standard Lambda Cold Dark Matter (\lcdm) cosmological model \citep{2006Sci...313..311B,Planck2016}. Mapping and interpreting the nearby cluster population is therefore essential both for understanding environmental effects on galaxies and for exploiting clusters as precision cosmological probes.

Traditional large-scale structure analyses rely on summary statistics, chiefly the two-point correlation function and power spectrum, to constrain cosmology \citep{2004ApJ...606..702T,2005MNRAS.362..505C}. While powerful under nearly Gaussian initial conditions, these statistics discard much of the non-Gaussian information encoded in the evolved morphology of the cosmic web, including higher-order correlations and phase information \citep{2010MNRAS.408.2163A,2013ApJ...762..115O,2022A&A...661A.146B}. Crucially, because summary statistics constrain only ensemble-averaged properties, they cannot recover the specific three-dimensional realization of the density field needed to link individual observed clusters to their simulated counterparts within a forward model.

Field-level Bayesian inference addresses this limitation by serving two complementary roles: (i) constraining cosmological parameters from the full information content of the observed galaxy distribution, and (ii) reconstructing a complete realization of the underlying dark-matter distribution with quantified uncertainties. The approach treats the initial conditions as parameters, forward models their nonlinear evolution to the present day, and constrains them with galaxy data to obtain a posterior ensemble. The \textit{Bayesian Origin Reconstruction from Galaxies} (\borg) algorithm \citep{Jasche2013}, and subsequent applications by the Aquila Consortium\footnote{\href{https://aquila-consortium.org/}{https://aquila-consortium.org/}}, have reconstructed the local density field and its dynamical history using this technique \citep[e.g.][]{Jasche2019,Stopyra2024_COLA}. Related Bayesian approaches include BARCODE \citep{Bos2016}, which reconstructs cosmic web environments from constrained realizations, and COSMIC BIRTH \citep{Kitaura2021}, an efficient framework for recovering the evolving cosmic web. Beyond Bayesian methods, complementary reconstruction techniques include least-action dynamical approaches such as the extended Fast Action Minimization method \citep{Sarpa2022}, which recovers galaxy trajectories and cosmic-web environments, and neural-network-based methods that infer dark-matter distributions from galaxy observables \citep{Veena2023}. Constrained-simulation projects such as the \textit{Constrained Local UniversE Simulations} (CLUES; \citealt{Gottloeber2010}), the \textit{Exploring the Local Universe with Reconstructed Initial Density Field} (ELUCID) simulations \citep{Wang2014,Wang2016}, the Local Universe Model \citep{Pfeifer2023}, and \slow \citep{Dolag2023} have produced halo and cluster catalogues conditioned on observational data, reproducing local structures and assessing their statistical properties.

Building on this foundation, \manticorelocal \citep{McAlpine2025} represents a state-of-the-art Bayesian reconstruction of the nearby Universe, combining high resolution, full nonlinear dynamics, and comprehensive posterior validation. Fitting a physical structure formation model to the \tmpp galaxy catalogue \citep{Lavaux2011,Carrick2015}, it produces a posterior ensemble of high-resolution constrained simulations—a ``digital twin'' of our cosmic neighbourhood—that accurately reproduces the observed large-scale structure. The reconstruction identifies high-significance counterparts for fourteen prominent galaxy clusters; including Virgo, Coma, and Perseus, each within one degree of its observed sky position, with inferred masses and redshifts in close agreement with observational estimates. In comparisons against the specific velocity field models tested by \citet{Stiskalek2025} (including linear theory, the \tmpp nonlinear reconstruction, and machine-learning approaches) the recovered peculiar velocity field achieves the highest Bayesian evidence across five independent datasets. Extensive validation confirms the local supervolume ($R < 200$ Mpc) is statistically consistent with \lcdm expectations, with no evidence for a large-scale underdensity.

Our primary goal for this study is to deliver a publicly available catalogue of cluster-scale objects in the local Universe derived from the \manticorelocal\ posterior. We term these objects \textit{associations}: groups of haloes across posterior realizations that correspond to the same inferred galaxy cluster, providing probabilistic counterparts to individual massive systems. While earlier constrained-simulation projects such as CLUES and ELUCID have also produced observationally anchored cluster catalogues enabling object-level comparisons, the key advance of the present work is the provision of a Bayesian posterior ensemble with fully quantified uncertainties. Unlike unconstrained \lcdm simulation catalogues (e.g., Millennium, \citealt{Springel2005}; Bolshoi-MDPL2, \citealt{Klypin2016}; IllustrisTNG, \citealt{Nelson2019}) that represent statistical populations without observational anchoring, and beyond single-realization constrained simulations that lack uncertainty estimates, our catalogue provides robust posterior distributions for the mass, position, velocity, and formation history of each association. This uncertainty quantification enables more rigorous comparisons: researchers can validate cluster properties against specific observed systems while properly accounting for reconstruction uncertainties, identify individual outliers that warrant targeted follow-up, and flag potential systematic biases in observational mass proxies on a case-by-case basis. The catalogue further supports hybrid analyses that combine observed quantities (e.g., X-ray profiles, lensing signals) with simulated properties (e.g., merger histories, dynamical states) for the same physical systems, enabling applications such as mass calibration studies, tests of hydrostatic equilibrium, and environmental quenching investigations. We validate the resulting catalogue through independent observables, including stacked thermal Sunyaev--Zel'dovich measurements and cross-matches with external X-ray catalogues, providing evidence that our inferred associations correspond to genuine massive structures. The catalogue and ancillary metrics are publicly available at \href{https://cosmictwin.org}{\texttt{cosmictwin.org}}.

The remainder of the paper is organized as follows. \cref{sect:method} describes the \manticorelocal data products and our procedure for extracting associations from the posterior ensemble. \cref{sect:results} presents the properties of the identified associations and their validation against independent datasets. \cref{sect:conclusions} summarizes our findings and outlines future applications.

\section{Methods}
\label{sect:method}

In this section we describe the construction and validation of a catalogue of massive structures, referred to as \textit{halo associations}. Our analysis begins from the \manticorelocal\ posterior suite of constrained $N$-body simulations, which provides a statistically consistent ensemble of realizations of the local Universe conditioned on galaxy observations. Within this ensemble we identify stable, recurrent structures—objects that appear across many posterior realizations at consistent locations and masses—and interpret these as probabilistic counterparts of galaxy clusters. The identification procedure is explicitly designed to prioritize robustness and reproducibility across the posterior rather than maximal completeness. We then describe the external observational data products used for validation, including \textit{Planck} Compton-$y$ maps and two complementary X-ray cluster catalogues, and finally outline the aperture-stacking methodology used to measure the thermal Sunyaev--Zel'dovich (tSZ) signal at the association positions.

\subsection{The \manticorelocal\ posterior suite}

This work leverages the \manticorelocal\ posterior suite, a collection of constrained cosmological simulations of the nearby Universe presented in \citet{McAlpine2025}. \manticorelocal\ employs field-level Bayesian inference to reconstruct the three-dimensional matter distribution of the local supervolume ($R \lesssim 200$~Mpc) from galaxy observations in the \tmpp\ catalogue. The inference framework builds upon the \textit{Bayesian Origin Reconstruction from Galaxies} (\borg) algorithm \citep{Jasche2013} and the \manticore\ model, incorporating refined galaxy-bias modelling, $\Lambda$CDM-conditioned priors, and a generalized Poisson likelihood. Together, these ingredients ensure statistical consistency with $\Lambda$CDM expectations while maintaining a high-fidelity match to the data.

The key modelling ingredients are as follows. The initial density field is represented on a $256^3$ grid with $\sim 3.9$~Mpc cell size and evolved forward using approximate gravity solvers to predict galaxy counts. A flexible, multi-parameter galaxy bias model allowing for variation across galaxy subsamples is jointly inferred with the initial density field, ensuring that bias--density degeneracies are marginalized over rather than fixed. The likelihood is a generalized Poisson distribution that accounts for over-dispersion in galaxy counts and incorporates survey selection functions. The inference is conditioned on $\Lambda$CDM priors, including Gaussian initial conditions and cosmological parameters consistent with Dark Energy Survey year three \citep[DES Y3,][]{DEScosmo} '3 × 2pt + All Ext.': $h=0.681$, $\Omega_\mathrm{m} = 0.306$, $\Omega_\mathrm{\Lambda} = 0.694$, $\Omega_\mathrm{b} = 0.0486$, $A_\mathrm{s} = 2.099 \times 10^{-9}$, and $n_\mathrm{s} = 0.967$. Small-scale modes beyond the inference grid resolution are unconstrained and are stochastically completed when generating posterior samples. As a result, large-scale structure is tightly constrained by the data, while halo-scale properties inherit additional scatter from the small-scale completion; this propagates into the positional and mass uncertainties of the associations identified in this work.

The \manticorelocal\ suite comprises 80 independent posterior resimulations, each representing a statistically plausible realization of the local Universe consistent with both the observed galaxy distribution and the assumed cosmological model. The original \citet{McAlpine2025} analysis used 50 realizations; we have since extended this to 80 to improve posterior sampling, and use the expanded suite throughout this work. These simulations are generated by sampling initial conditions from the inferred posterior and evolving them with the \swift\ $N$-body code \citep{Schaller2024} in a periodic domain of side length $L=1000$~Mpc. Only the inner $\sim 200$~Mpc are directly constrained by observational data with the highest signal-to-noise; the outer regions follow unconstrained $\Lambda$CDM priors. Each realization contains $1024^3$ dark-matter particles with mass $3.7\times 10^{10}\,{\rm M_\odot}$, such that the lowest-mass systems considered in this study ($M_{200}\sim 7.5\times 10^{13}\,{\rm M_\odot}$) are resolved with $\sim 2{,}000$ particles. The statistical fidelity of the \manticorelocal\ reconstruction (including consistency with $\Lambda$CDM power spectra, halo mass functions, and velocity fields) has been demonstrated in \citet{McAlpine2025}.

Dark-matter haloes are identified in each posterior realization using the \texttt{HBT-HERONS} subhalo finder \citep{2025arXiv250206932F}, which traces bound particle histories to robustly track haloes and their substructure. Halo properties are computed with the \texttt{SOAP} analysis pipeline \citep{2025arXiv250722669M}, which measures spherical-overdensity masses $M_{200,\mathrm{crit}}$ and $M_{500,\mathrm{crit}}$. Here $M_{200}$ denotes the mass enclosed within a sphere of mean density 200 times the critical density; we drop the ``crit'' subscript throughout for brevity, and similarly for $M_{500}$. We restrict the analysis to central haloes at $z=0$, defined as the most massive subhalo within each halo by construction. All halo positions and velocities are expressed in an observer-centric equatorial coordinate system, with the fiducial observer placed at $(500,500,500)$~Mpc within the simulation volume, enabling direct comparisons to observed local clusters (e.g.\ Virgo, Coma, Perseus).

\begin{figure}
    \includegraphics[width=\columnwidth]{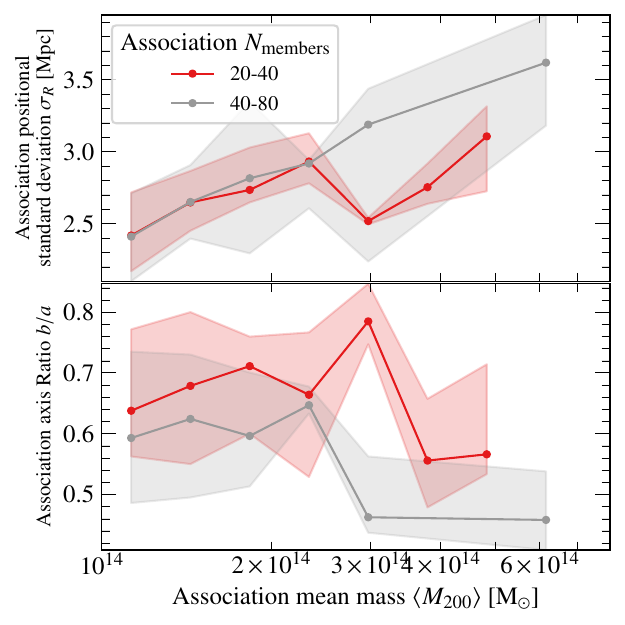}
    \caption{Median association properties as a function of mean mass $\langle M_{200} \rangle$, shown separately for associations with $20$--$40$ (red) and $40$--$80$ (grey) member haloes.
\textit{Top}: Three-dimensional positional scatter $\sigma_R$.
\textit{Bottom}: Axis ratio $b/a$, where $b/a = 1$ indicates spherical symmetry and lower values indicate elongation.
Shaded regions show the 25th--75th percentiles.
Associations exhibit comparable spatial extents up to $\langle M_{200}\rangle \simeq 3\times10^{14}\,{\rm M_\odot}$, while at higher masses systems with more members become increasingly extended.
These higher-mass associations are also more asymmetric (lower $b/a$), and the growth of this asymmetry is reflected in the increase of the three-dimensional positional scatter $\sigma_R$.
    }
    \label{fig:position_std_vs_mass}
\end{figure}

\begin{figure}
    \includegraphics[width=\columnwidth]{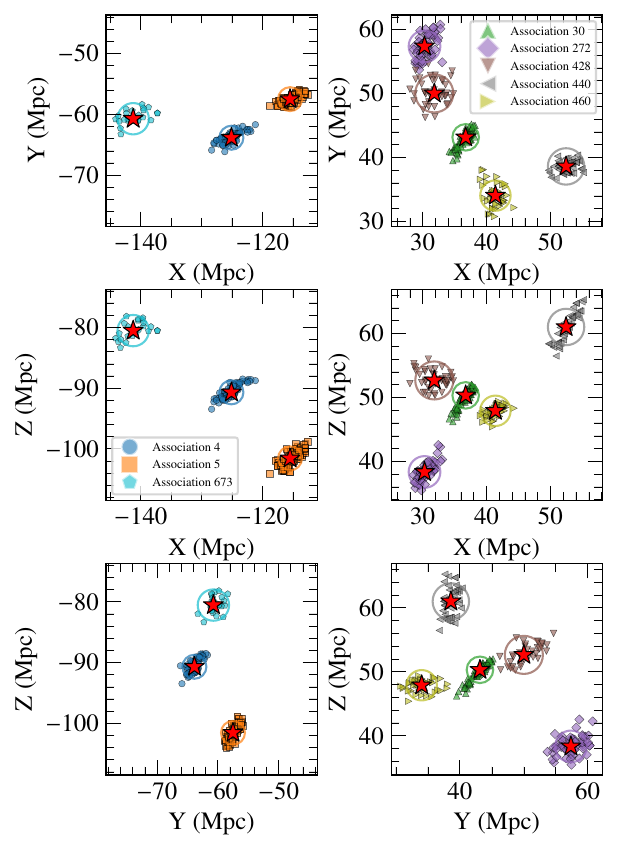}
    \caption{Examples of identified halo associations in three orthogonal projections at the fiducial DBSCAN parameters ($\epsilon = 2.75$~Mpc, $N_{\mathrm{min}} = 10$). Distinct associations are shown with different colours and symbols; circles mark the $1\sigma$ positional scatter of member haloes across the posterior. The examples illustrate that the clustering procedure recovers spatially coherent structures. A systematic analysis of parameter sensitivity is provided in \cref{app:dbscan}.}
    \label{fig:example_halo_groupings}
\end{figure}

\subsection{Identification and robustness of halo associations}
\label{subsec:assoc_method}

We seek objects that recur across the \manticorelocal\ posterior in consistent locations and masses. An \textit{association} is defined as a set of $z=0$ central haloes from different posterior realizations that represent the same inferred galaxy cluster, i.e.\ the cluster-scale structure that the Bayesian reconstruction places at a given location across multiple posterior samples. Associations are identified by clustering haloes in Eulerian space, with each posterior realization contributing at most one \textit{member halo} to any given association (this term does not imply a satellite or subhalo; it simply denotes the single representative halo from that realization).

The input sample comprises all $z=0$ central haloes with $M_{200} \ge 7.5 \times 10^{13}\,{\rm M_\odot}$ from each of the 80 realizations across the full reconstructed volume. The mass threshold is set below our target of $\langle M_{200} \rangle \ge 10^{14}\,{\rm M_\odot}$ for the final catalogue because individual member haloes within an association can scatter below the mean mass due to posterior uncertainty, and we wish to retain these members for accurate characterization of the posterior distribution.

Associations are identified using the Density-Based Spatial Clustering of Applications with Noise algorithm (DBSCAN; \citealt{10.5555/3001460.3001507}) applied in three-dimensional comoving Cartesian coordinates. DBSCAN is a density-based clustering algorithm controlled by two parameters: the neighbourhood radius $\epsilon$, and the minimum number of neighbours $N_{\mathrm{min}}$ required within $\epsilon$ to seed a cluster. Points with at least $N_{\mathrm{min}}$ neighbours are classified as core points; clusters grow by recursively linking core points and their neighbours into connected components. Points that cannot be reached from any dense region are labelled as noise. We adopt fiducial parameters $\epsilon = 2.75$~Mpc and $N_{\mathrm{min}} = 10$, derived from a systematic sensitivity analysis described in \cref{app:dbscan}. Each realization may contribute at most one member halo to a given association; when DBSCAN initially links multiple haloes from the same realization, only the halo closest to the current association centroid is retained. The frequency with which this one-per-realization constraint is invoked defines the \textit{ambiguity rate} of that association, which serves as a diagnostic for over-merging.

From the raw DBSCAN output we construct the final catalogue by applying quality cuts based on ambiguity rate and membership. The \textit{fiducial} sample requires an ambiguity rate below 5\% to ensure clean initial assignments, and at least 20 member haloes ($N_{\rm members} \ge 20$) to ensure adequate sampling of the posterior for reliable estimation of summary statistics such as mean position and mass. A \textit{strict} sample with additional constraints on positional and mass scatter is also provided for applications requiring the most robust predictions. The full selection criteria and their motivation are detailed in \cref{app:dbscan} and summarized in \cref{tab:selection_cuts}.

The catalogue is robust against chance alignments: applying the identical pipeline to control simulations with randomized observer positions yields zero associations at the fiducial parameters, confirming that the detected structures reflect genuine spatial correlations imposed by observational constraints rather than statistical fluctuations. The sensitivity of results to parameter choices is quantified in \cref{app:dbscan}, where we demonstrate that the fiducial parameters are derived from physically motivated diagnostics, the ambiguity rate constrains $\epsilon$ to avoid over-merging, while a membership-ratio diagnostic constrains $N_{\mathrm{min}}$ to filter noise without discarding real associations.

For each association we compute posterior summaries of mass, positional scatter, and shape. As shown in \cref{fig:position_std_vs_mass}, associations exhibit comparable spatial extents up to $\langle M_{200}\rangle \simeq 3\times10^{14},{\rm M_\odot}$, beyond which systems with a larger number of members become systematically more extended. In the bottom panel, associations with more members are typically more asymmetric (lower $b/a$), with this asymmetry increasing above $\sim 3\times10^{14},{\rm M_\odot}$; this enhanced elongation drives the corresponding increase in the three-dimensional positional scatter $\sigma_R$ at high masses. \cref{fig:example_halo_groupings} shows representative examples of identified associations in three orthogonal projections, illustrating the spatial coherence of the recovered structures.

\subsection{External data products}

We use both Sunyaev--Zel'dovich and X-ray datasets as independent observational benchmarks. For the thermal SZ signal we adopt the \textit{Planck} PR4 Compton-$y$ map produced with the needlet internal linear combination (NILC) method \citep{2023MNRAS.526.5682C}. The PR4 processing offers improved noise and foreground mitigation relative to earlier releases. The map is delivered at $N_{\rm side}=2048$ ($\sim 10'$ effective beam); we apply the official PR4 masks (Galactic, point source, and quality) to exclude contaminated regions. This dataset is used in stacking analyses (see \cref{sect:stacking_procedure,sect:tsz_stacking}) to test that our associations correspond to real SZ signals on the sky.

For X-ray clusters we rely on two complementary catalogues. The MCXC-II meta-catalogue \citep{2024A&A...688A.187S} provides a homogenized compilation of ROSAT-detected clusters with masses standardized to $M_{500}$. The eROSITA eRASS1 catalogue \citep{2024A&A...685A.106B} contains modern detections with masses calibrated against weak lensing \citep{2024A&A...689A.298G}, defining an updated X-ray mass scale. These catalogues serve two roles: (i) to confirm that our stacking pipeline recovers SZ signals at the sky locations of known clusters, and (ii) to provide an external reference for comparing our halo mass estimates (see \cref{sec:mass_ratio_external}). These products are used only for validation; they were not inputs to the \manticorelocal\ inference.

\subsection{Compton-$y$ aperture-stacking procedure}
\label{sect:stacking_procedure}

To test whether posterior associations correspond to genuine massive structures, we perform a thermal Sunyaev--Zel'dovich (tSZ) stacking analysis using the \textit{Planck} PR4 NILC Compton-$y$ map. Each association is treated as a single halo centred on its posterior mean sky position. The tSZ signal is measured using compensated aperture photometry with an inner aperture of radius $1.5\,R_{500}$ and an outer background annulus extending from $1.5$ to $5.0\,R_{500}$. The primary observable is the aperture contrast
\begin{equation}
\Delta y \equiv \langle y_{\rm inner} \rangle - \langle y_{\rm outer} \rangle ,
\end{equation}
where averages are taken over unmasked pixels. This compensated design suppresses large-scale fluctuations and foregrounds while isolating the local cluster signal. Associations with less than $80\%$ unmasked coverage in either the inner or outer region are discarded. All associations contribute with equal weight.

We apply the official PR4 Galactic, point-source, and quality masks and otherwise use the NILC map at its native $\sim 10'$ resolution, without additional filtering. Large-scale residuals are further reduced by subtracting the median signal in a $5^\circ$--$7^\circ$ annulus around each association prior to photometry, following standard practice.

Detection significance is derived from the ensemble of individual aperture measurements rather than from the stacked images. We compute
\begin{equation}
{\rm S/N} = \frac{\langle \Delta y \rangle}{\sigma_{\rm boot}} ,
\end{equation}
where $\langle \Delta y \rangle$ is the mean aperture contrast across all associations in the sample, and $\sigma_{\rm boot}$ is the bootstrap standard error estimated from $N_{\rm boot}=100$ resamplings. The bootstrap accounts for both measurement uncertainty and sample variance. This statistic quantifies the significance of the ensemble tSZ detection; it is not tied to the amplitude of the stacked radial profiles shown for visualization.

As a null test, we repeat the entire analysis on random sky positions drawn to match the $R_{500}$ distribution and mask-coverage properties of the \manticorelocal\ sample, using the same aperture photometry and bootstrap procedure. These null stacks yield results consistent with zero, confirming that the measurement is not driven by map artefacts or masking systematics.

Finally, we construct stacked radial profiles by averaging the background-subtracted Compton-$y$ signal in concentric annuli. Profiles are shown in units of $r/R_{500}$ for scale-free comparison. These profiles serve as a qualitative diagnostic of the spatial concentration of the signal and are not used in the significance calculation.

\section{Results}
\label{sect:results}

We now turn to the main scientific outcomes from the \textsc{Manticore-Local} posterior cluster catalogue.
Unless otherwise specified, all analyses use the fiducial catalogue of 401 posterior associations defined in \cref{app:dbscan}, which requires $\langle M_{200} \rangle \geq 10^{14}\,{\rm M_\odot}$, ambiguity rates below 5\%, and at least 20 member haloes across the posterior ensemble.
We begin with a census of these robust halo associations and a check for independent signatures in \textit{Planck} tSZ maps.
We then examine how present-day constraints shape the inferred evolutionary pathways of these systems, and conclude with a comparison to external X-ray mass estimates.
Together, these analyses demonstrate both the reliability of the catalogue and the new types of cluster-level science enabled by field-level inference.

\subsection{Census of posterior halo associations}
\label{sect:census}

\begin{figure}
    \centering
    \includegraphics[width=\columnwidth]{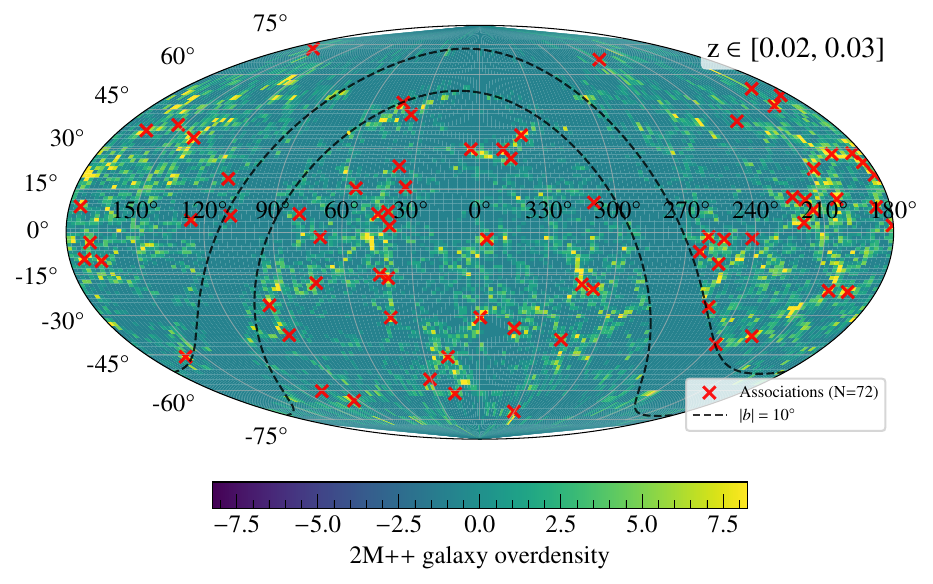}
    \caption{Sky distribution of posterior halo associations overlaid on the \tmpp\ galaxy overdensity field $\delta = (N - \langle N \rangle)/\langle N \rangle$ in a Mollweide projection within the redshift range $z \in [0.02, 0.03]$. Red crosses mark the 72 associations within the redshift slice. Dashed lines indicate the Zone of Avoidance at Galactic latitude $|b| = 10^\circ$, where survey completeness is degraded by stellar contamination and dust extinction. The associations preferentially trace overdense regions (yellow), demonstrating that they correspond to genuine large-scale structure rather than artefacts of the identification procedure.}
    \label{fig:spatial_distribution}
\end{figure}

\begin{figure}
    \centering
    \includegraphics[width=\columnwidth]{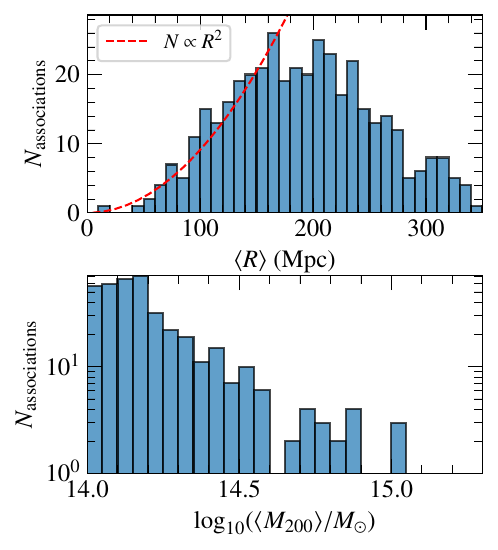}
    \caption{Distance and mass distributions of the 401 posterior halo associations in the fiducial catalogue. \textit{Top}: distribution of mean observer distances. The red dashed line indicates the expected $N \propto R^2$ scaling for a uniform spatial distribution in Euclidean space. The distribution broadly follows this scaling to $\approx$150~Mpc before declining at larger distances, reflecting the degradation of \tmpp\ constraints with distance. \textit{Bottom}: distribution of mean masses $\langle M_{200} \rangle$, shown on a logarithmic scale. The sample is dominated by systems near the $10^{14}$~M$_\odot$ threshold, with a steeply declining tail toward higher masses consistent with the halo mass function.}
    \label{fig:distance_distribution}
\end{figure}

\begin{figure*}
    \includegraphics[width=\textwidth]{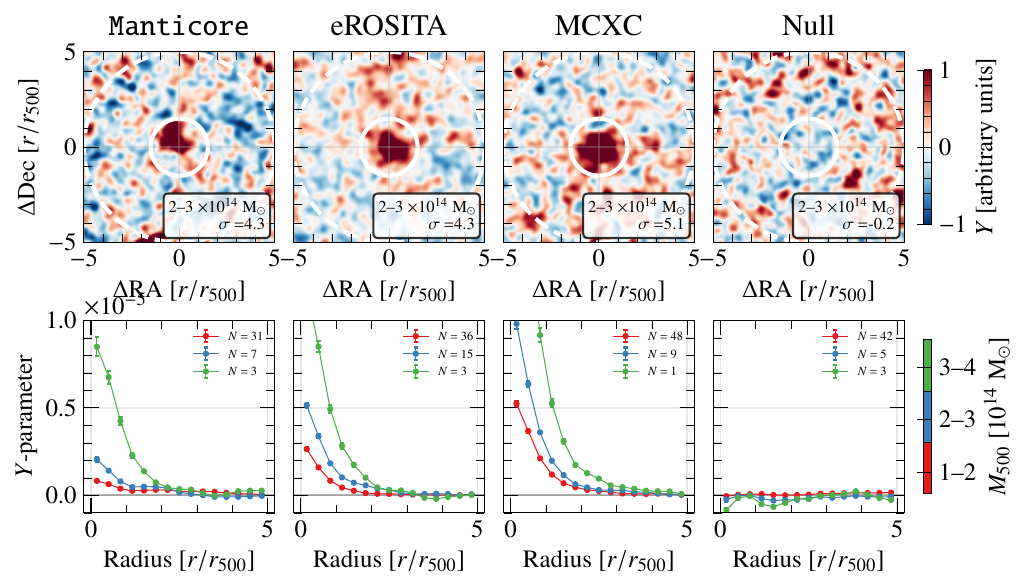}
    \caption{\textit{Top row:} Stacked \textit{Planck} PR4 NILC Compton-$y$ patches for systems with $2\times10^{14} \leq M_{500}/M_\odot < 3\times10^{14}$ and $z \leq 0.06$, drawn from the \textsc{Manticore-Local} posterior associations (fiducial sample), the eROSITA eRASS1 catalogue, the MCXC-II meta-catalogue, and a null test using random sky positions. White circles indicate the inner aperture radius ($1.5\,R_{500}$); the outer background annulus extends to $5\,R_{500}$. Detection significances ($\sigma$) are computed from individual aperture measurements prior to stacking. \textit{Bottom row:} Corresponding stacked radial $y$-profiles in units of $r/R_{500}$ with bootstrap uncertainties for three mass bins ($N$ indicates the number of clusters per bin). \textsc{Manticore-Local} produces clear, spatially localized tSZ decrements comparable to those obtained from external catalogues, while the null test shows no signal. Results using the strict sample are visually indistinguishable and are omitted for clarity.}
    \label{fig:stacked_patches}
\end{figure*}

\begin{figure*}
    \includegraphics[width=\textwidth]{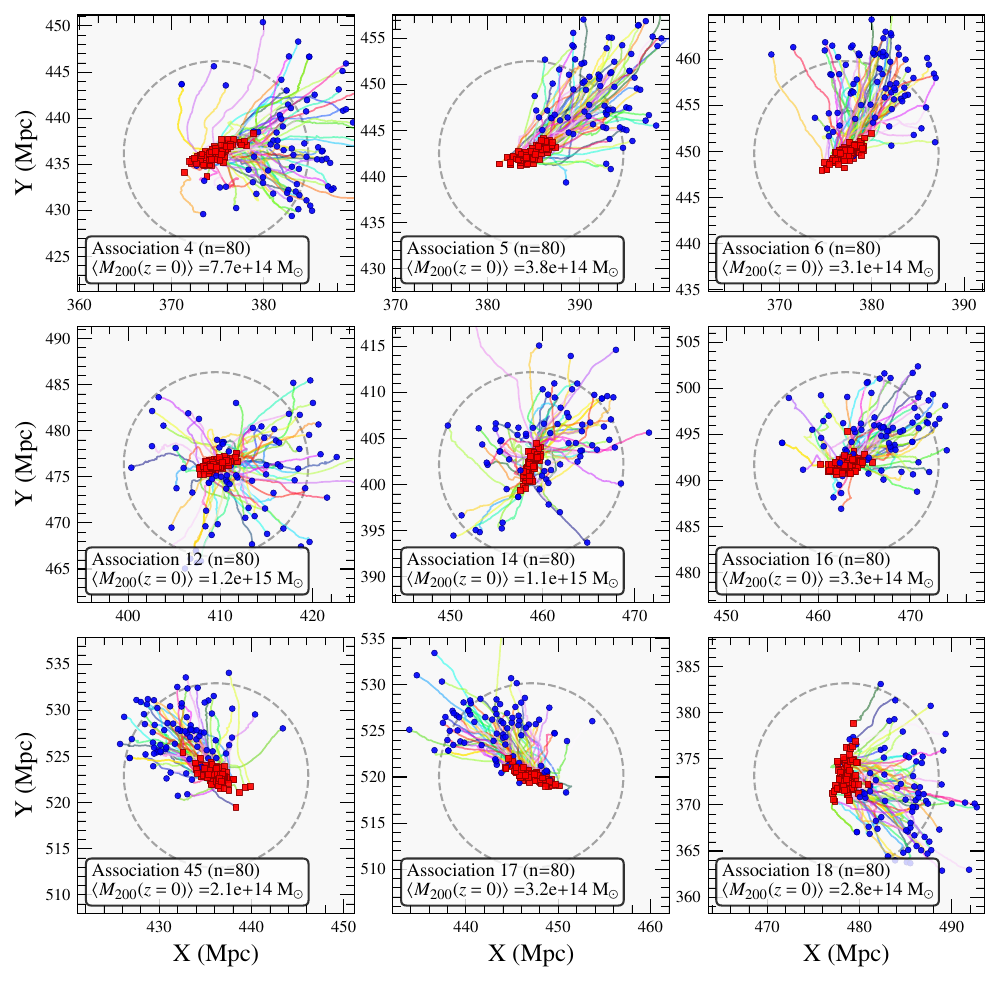}
    \caption{
    Evolutionary tracks of nine representative posterior associations with the largest number of member haloes. Each line traces the main progenitor branch of a halo from $z=5$ (blue circles) to $z=0$ (red squares) in comoving coordinates. Colours distinguish different members. A dashed grey ring of radius $10\,\mathrm{Mpc}$ (comoving) is drawn around the $z=0$ centroid for scale. The association ID, number of members $n$, and mean $\langle M_{200}(z=0) \rangle$ are annotated. These examples illustrate that posterior associations do not emerge from random initial conditions, but more often from localized regions of Lagrangian space with diverse, yet coherent, assembly pathways.}
    \label{fig:posterior_halo_histories}
\end{figure*}

Our clustering analysis yields a fiducial catalogue of 401 posterior halo associations with $\langle M_{200} \rangle \geq 10^{14}\,{\rm M_\odot}$, representing robustly constrained massive halos in the local Universe that are consistently recovered across the 80 posterior realizations.

\Cref{fig:spatial_distribution} shows the sky distribution of associations overlaid on the \tmpp\ galaxy overdensity field within the redshift range $z \in [0.02, 0.03]$. The associations preferentially coincide with overdense regions in the galaxy distribution, confirming that they trace genuine large-scale structure rather than artefacts of the clustering algorithm. The reduced density of associations near the Galactic plane (marked by dashed lines at $|b| = 10^\circ$) reflects the lower completeness of the \tmpp\ survey in this region due to stellar contamination and dust extinction. Nevertheless, a small number of associations are recovered within the Zone of Avoidance, consistent with structures that remain detectable despite degraded constraints.

The radial distribution of mean observer distances for the posterior halo associations is shown in \cref{fig:distance_distribution}. At distances $\lesssim 150$~Mpc from the observer, the distribution broadly follows the expected $N \propto R^2$ scaling for a uniform population in Euclidean space. Beyond this scale, the counts decline as observational constraints become noisier with increasing distance, causing the average cluster significance to decrease and fewer associations to meet our membership criteria. This is consistent with the signal-to-noise decline beyond $R \approx 200$~Mpc demonstrated in \citet{McAlpine2025}. This reflects a fundamental limitation of the catalogue: completeness degrades with distance as the underlying \tmpp\ constraints weaken. Rather than include poorly constrained systems, we retain only associations consistently recovered across the posterior, accepting reduced completeness at large distances in favour of reliability. The mass distribution (lower panel of \cref{fig:distance_distribution}) shows that the sample is dominated by systems near the $10^{14}$~M$_\odot$ threshold, declining steeply toward higher masses as expected from the halo mass function.

\subsection{Independent validation through tSZ detection}
\label{sect:tsz_stacking}

In \citet{McAlpine2025}, the accuracy of the \manticorelocal reconstruction was validated through detailed object-by-object comparisons of 14 prominent local clusters including Virgo, Coma, Perseus, and Norma. That analysis demonstrated: (i) positional accuracy exceeding the inference grid resolution, with median angular separations of $1$--$2^\circ$ compared to the $\sim 3$--$5^\circ$ voxel scale; (ii) mass posteriors consistent with multi-wavelength observational estimates for all 14 clusters, with coefficients of variation of $0.2$--$0.35$ for massive systems; and (iii) a reconstructed velocity field achieving the highest Bayesian evidence across five independent peculiar velocity catalogues. Limitations include increased positional uncertainty for nearby clusters with large peculiar velocities (e.g.\ Virgo) and one outlier (Hercules~A2147) with a slightly underestimated distance. Here we perform a complementary, population-level consistency check by testing whether the ensemble of posterior halo associations coincides with hot gas structures in the real Universe. We emphasize that this analysis is not intended to derive $Y$--$M$ scaling relations or perform a full cosmological tSZ study; rather, it serves as an independent sanity check that our associations are located where real clusters exist. We use stacking measurements of the thermal Sunyaev--Zel'dovich (tSZ) signal in the \textit{Planck} PR4 NILC Compton-$y$ map (see \Cref{sect:stacking_procedure} for methodological details). The tSZ field provides an entirely independent observable: it traces thermal pressure in the intracluster medium rather than the galaxy distribution that informs \manticorelocal. A coherent decrement in the stacked signal therefore provides supporting evidence that the inferred halo associations are spatially coincident with genuine massive structures.

We use two established X-ray cluster catalogues---the MCXC-II meta-catalogue \citep{2024A&A...688A.187S} and the eROSITA eRASS1 sample \citep{2024A&A...685A.106B}---primarily as methodological benchmarks to validate our stacking pipeline. These catalogues are not used as ground truth, nor do we attempt to match absolute Compton-$y$ amplitudes between catalogues. Their role is to demonstrate that our pipeline recovers known tSZ signals with the expected behaviour. For this validation exercise, all samples are restricted to $1\times10^{14} \leq M_{500}/M_\odot < 4\times10^{14}$ and $z \leq 0.06$, matching the depth of the \manticorelocal sample. We additionally require $\geq 80\%$ unmasked aperture coverage in the \textit{Planck} $y$-map and restrict to Galactic latitudes $|b|\in[10^\circ,90^\circ]$ to mitigate foregrounds and projection distortions. These cuts define the validation subset used for stacking and do not affect the completeness of the \manticorelocal catalogue itself. After these selections, 71 of the 401 \manticorelocal posterior associations remain in the validation sample.

As shown in \cref{fig:stacked_patches}, \manticorelocal produces a clear, centrally concentrated tSZ decrement at the expected cluster positions. In the mass bin $2\times10^{14} \leq M_{500}/M_\odot < 3\times10^{14}$, we obtain a detection significance of $\sigma = 4.3$, where $\sigma$ denotes the signal-to-noise ratio of the ensemble aperture measurements (computed before stacking). This is far above the level expected from random alignments: the null test yields $\sigma \approx 0$. The external catalogues show similarly significant detections ($\sigma = 4.3$ for eROSITA and $\sigma = 5.1$ for MCXC-II), confirming that the stacking pipeline behaves consistently across all three cluster samples. These values are not intended for direct amplitude comparison between catalogues. The MCXC-II sample is based on hydrostatic X-ray masses, while eROSITA masses rely on weak-lensing calibrated scaling relations \citep{2024A&A...685A.106B}, and the resulting mass bins therefore contain different effective populations. X-ray selection also favours systems with the strongest tSZ signals, whereas \manticorelocal is not selected on gas observables. Differences in the detailed profile shapes should therefore not be interpreted as discrepancies. The reduced central amplitude of the \manticorelocal stacked profile is naturally explained by the $\sim$2--3~Mpc positional uncertainties within the posterior ensemble (\cref{fig:position_std_vs_mass}), which dilute the stacked signal.

Most importantly, \manticorelocal reproduces the expected trend with halo mass: higher-mass bins show increasingly strong tSZ decrements, consistent with the established $M_{500}$--$Y$ scaling relation. This scaling is recovered despite the positional uncertainties and despite no tSZ information being used in the reconstruction, indicating that the posterior associations track the true underlying halo population. Repeating the analysis with the strict catalogue selection yields visually indistinguishable stacked profiles and consistent detection significances, confirming that the results are robust to the choice of selection criteria.

The key result is qualitative: \manticorelocal associations produce statistically significant, spatially concentrated tSZ decrements where expected, providing evidence that they trace genuine massive structures in the real Universe. The external catalogues confirm that our stacking pipeline behaves consistently on known cluster samples, while the null test demonstrates that the detections are not driven by map artefacts. This ensemble-level consistency check complements the object-level comparisons of \citet{McAlpine2025}, providing independent support for the spatial fidelity of the \manticorelocal reconstruction. We emphasize that this test validates the ensemble on average; it does not confirm individual associations, nor does it test mass accuracy or sample purity. A full quantitative cosmological analysis would require careful treatment of selection effects, mass calibration, positional uncertainties, and their propagation into the error budget, which lie beyond the scope of this consistency check.

\subsection{Restricted evolutionary pathways of posterior associations}
\label{sect:cluster_evolution}

Having validated the \textsc{Manticore-Local} catalogue against independent hot-gas measurements (\cref{sect:tsz_stacking}), we now examine what the inference reveals about the \textit{histories} of the posterior associations. The central question is whether observational constraints on the galaxy distribution at $z=0$ alone, without direct velocity measurements or high-redshift data, suffice to restrict the range of possible formation pathways that led to these present-day structures.

The methodology is as follows. Each posterior realization is a self-consistent N-body simulation with its own merger tree structure. For a given posterior association, we extract the main progenitor branch from the merger tree of each member halo within its respective realization, tracing positions backward from $z=0$ to $z=5$. The ensemble of trajectories shown in subsequent figures therefore represents the range of formation pathways consistent with the $z=0$ galaxy data. These are individual trajectories from individual realizations, not averaged or merged distributions.

\begin{figure}
    \includegraphics[width=\columnwidth]{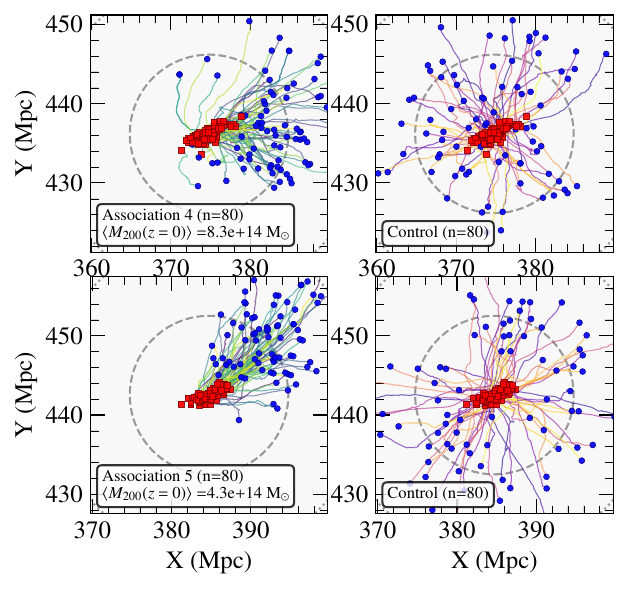}
    \caption{
    Evolutionary trajectories for two posterior associations (left) and their matched controls (right). Red squares mark $z=0$ positions, blue circles mark $z=5$, and lines trace progenitor branches. Dashed circles indicate $10\,\mathrm{Mpc}$ radii. Despite identical final positions by construction, the constrained haloes exhibit more organized infall patterns with preferred directions, whereas controls are much more dispersed. This demonstrates that observational constraints at $z=0$ meaningfully reduce the range of viable evolutionary pathways.}
    \label{fig:control_trace_example}
\end{figure}

\begin{figure*}
    \includegraphics[width=\textwidth]{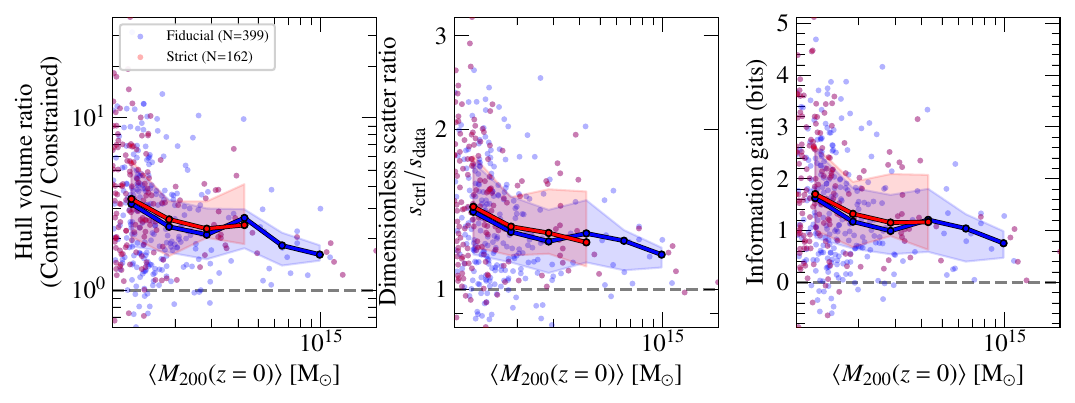}
    \caption{
    Population-level comparison of Lagrangian localization metrics for posterior associations with $\langle M_{200}(z=0) \rangle > 10^{14} M_\odot$, shown separately for the fiducial (blue; $N=399$) and strict (red; $N=162$) catalogue selections defined in \cref{app:dbscan}.
    \textit{Left:} Hull volume ratio $\mathcal{R}_V$, showing that control haloes occupy typically $2$--$5$ times larger initial volumes.
    \textit{Middle:} Scatter ratio $\mathcal{R}_s$, indicating $20$--$50\%$ larger dispersions in controls.
    \textit{Right:} Information gain $I$, with median $I \approx 1.5$ bits, corresponding to a factor of $\sim 3$ reduction in configuration-space volume.
    Small points show individual associations, large points the medians per mass bin, and shaded bands the interquartile ranges. Horizontal dashed lines mark the null expectation ($\mathcal{R}=1$, $I=0$).
    Both selections yield consistent results, with all three diagnostics indicating that posterior associations are more localized than matched controls. The strict sample shows marginally stronger localization, suggesting that associations with tighter $z=0$ constraints also tend to have better-constrained formation pathways.}
    \label{fig:trace_metric_compare}
\end{figure*}

\Cref{fig:posterior_halo_histories} provides a qualitative view for nine of the richest associations. Member haloes are traced along their main progenitor branches back to $z=5$. The resulting tracks show that systems typically originate from well-defined, spatially compact regions of Lagrangian space, with trajectories exhibiting coherent infall along preferred directions---that is, the ensemble of member trajectories shows correlated spatial structure rather than isotropic scatter. Some evolve from compact, near-isotropic patches, others from elongated filaments, and others from more dispersed initial configurations. In some cases haloes migrate only a few Mpc before assembling, while in others they travel more than $10\,\mathrm{Mpc}$. This variety is consistent with the diversity of cluster formation pathways predicted in $\Lambda$CDM \citep[e.g.][]{Bond1996,2014MNRAS.443.1090F,2018MNRAS.473.1195L}. Crucially, however, the posterior solutions do not explore this diversity randomly: they occupy restricted, anisotropic regions of initial configuration space. This localization indicates that the inference meaningfully restricts the range of plausible formation pathways, even though the input data used in the \manticorelocal inference were collected only at $z=0$. However, this should not be interpreted as uniquely determining formation histories: in nonlinear environments, multiple posterior-consistent initial configurations can still lead to similar $z=0$ structures while differing in their detailed assembly sequences.

To quantify this effect and isolate the impact of observational constraints, we compare each posterior association to "control" haloes drawn from unconstrained $\Lambda$CDM simulations with identical resolution and cosmology. For each constrained association, we select control haloes matched in two key properties: (i) final mass $M_{200}(z=0)$ within $\pm 0.1$ dex, and (ii) comoving travel distance between $z=5$ and $z=0$ within 20\%. These criteria ensure that constrained and control haloes have comparable bulk properties, they are equally massive and have traversed similar distances during their evolution, while leaving the spatial distribution and morphology of their progenitors entirely unconstrained. We note that this matching does not control for large-scale environment (e.g., proximity to filaments, local tidal anisotropy), which is known to influence the coherence and anisotropy of halo assembly \citep[e.g.][]{2007MNRAS.375..489H,2018MNRAS.473.1195L}. A more controlled comparison would additionally match on environmental properties; however, disentangling environmental selection from data constraints would require a more elaborate framework beyond the scope of this analysis. The present comparison therefore demonstrates the combined effect of observational constraints and any correlated environmental information encoded in the $z=0$ galaxy distribution. To enable direct spatial comparison, we translate each control halo so that its $z=0$ position coincides with the centroid of the corresponding constrained association.

\Cref{fig:control_trace_example} shows two representative examples. By construction, the final $z=0$ positions coincide, but the evolutionary tracks differ markedly. Constrained haloes exhibit coherent infall along preferred directions, with progenitors clustered in compact regions of Lagrangian space. In contrast, controls are scattered more isotropically across a much larger volume, despite having identical masses and travel distances. This difference directly demonstrates that present-day galaxy data restrict the allowed assembly pathways.

We quantify this localization using three complementary metrics applied to the spatial distribution of progenitors at $z=5$:
\begin{itemize}
    \item \textbf{Convex hull volume ratio} ($\mathcal{R}_V$): the ratio of the volumes enclosed by the convex hulls of control and constrained progenitors. Larger values indicate that controls occupy a proportionally larger initial volume.
    \item \textbf{Dimensionless scatter ratio} ($\mathcal{R}_s$): the ratio of root-mean-square dispersions (normalized by the mean distance from centroid) between control and constrained progenitors. Values above unity indicate greater spatial dispersion in controls.
    \item \textbf{Information gain} ($I$): derived from the ratio of covariance determinants, this metric quantifies the reduction in configuration-space volume in units of bits. An information gain of $I$ bits corresponds to a factor of $2^I$ reduction in effective volume.
\end{itemize}
All three metrics are constructed such that larger values indicate stronger localization of constrained progenitors relative to their matched controls.

\Cref{fig:trace_metric_compare} shows results for associations with $\langle M_{200}(z=0) \rangle > 10^{14} M_\odot$ and at least 80\% control-matching success, comparing the fiducial ($N=399$) and strict ($N=162$) catalogue selections. The localization effect is systematic and substantial across both samples. Median values for the fiducial sample are $\mathcal{R}_V \approx 3$, $\mathcal{R}_s \approx 1.3$, and $I \approx 1.5$ bits, corresponding to typical two- to four-fold reductions in the allowed configuration-space volume of progenitors. The strict sample shows consistent but marginally stronger localization, with slightly elevated median values across all three metrics. This mild enhancement is expected: associations that pass the strict selection criteria---requiring tighter spatial and mass scatter at $z=0$---represent systems where the posterior is more tightly constrained, and this improved constraint quality appears to propagate to their inferred formation pathways. However, the difference between selections is modest, and both samples support the same qualitative conclusion. Individual associations span a range of localization strengths, from strongly constrained systems such as Association~5 ($\mathcal{R}_V=6.6$, $I=3.2$ bits) to more mildly constrained cases like Association~4 ($\mathcal{R}_V=2.0$, $I=1.3$ bits), both shown in \cref{fig:control_trace_example}. Importantly, this localization shows only weak mass dependence over the range $10^{14}$--$10^{15} M_\odot$, indicating that the effect is not simply driven by the most massive, rarest systems, but applies broadly across the cluster population.

These results demonstrate that \textsc{Manticore-Local} not only recovers present-day cluster analogues, but also systematically reduces the configuration-space volume of allowed progenitor states relative to unconstrained $\Lambda$CDM. By propagating present-day galaxy data backward through the gravitationally coupled forward model in the inference, we achieve a systematic restriction of progenitor configurations---reducing the space of possibilities by typical factors of two to five---despite having imposed constraints only at $z=0$. We emphasize that this localization reflects information gain relative to the prior, not a validation of pathway accuracy: we lack direct observational constraints on these high-redshift progenitor configurations, and the inferred trajectories should therefore be interpreted as constrained realizations consistent with the data rather than definitive formation histories. Nevertheless, the internal consistency of the posterior ensemble and the agreement with $z=0$ validation tests (\cref{sect:tsz_stacking}) support the physical plausibility of these inferred pathways, confirming that field-level inference can provide informative constraints on the range of plausible formation histories for individual massive structures.

\begin{figure}
    \includegraphics[width=\columnwidth]{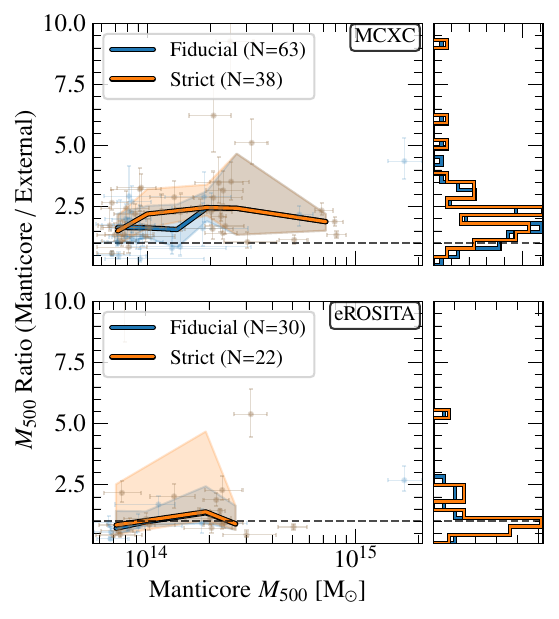}
    \caption{
    Comparison of \textsc{Manticore-Local} $M_{500}$ masses to external X-ray estimates from the MCXC-II meta-catalogue \citep[][top]{2024A&A...688A.187S} and the eROSITA eRASS1 catalogue \citep[][bottom]{2024A&A...685A.106B}.
    \textit{Left panels}: Individual cluster ratios as a function of \manticorelocal mass, with vertical error bars representing the 25th--75th percentile range of the posterior mass distribution. Shaded bands show binned medians with interquartile ranges.
    \textit{Right panels}: Normalized histograms of the ratio distributions.
    Blue points show the fiducial sample; orange points show the strict selection (\cref{app:dbscan}). Sample sizes indicate the number of associations with successful cross-matches after applying the criteria in \cref{sec:mass_ratio_external}: $N=63$ (fiducial) and $N=38$ (strict) for MCXC; $N=30$ and $N=22$ for eROSITA. The two selections yield consistent results, indicating that the mass ratios are robust to catalogue definition.
    \textsc{Manticore-Local} masses are systematically higher than MCXC-II values (median ratio $\sim 1.7$), while eROSITA shows a median ratio close to unity. The dashed horizontal line marks equality.}
    \label{fig:mass_ratio_external}
\end{figure}

\subsection{Comparison to external cluster mass estimates}
\label{sec:mass_ratio_external}

We conclude the results section by demonstrating a capability enabled by constrained catalogues: direct, object-by-object comparison of posterior masses to external measurements. Unlike traditional simulation-based studies that must rely on ensemble-averaged scaling relations to connect observations to theory, a digital twin of the local Universe allows us to compare individual observed clusters to their specific simulated counterparts. This enables systematic trends and outliers to be identified directly, without requiring statistical corrections for ensemble completeness or sample selection—each discrepancy can be traced to a specific physical system.

Here we focus on two independent X-ray--derived datasets: the MCXC-II compilation \citep{2024A&A...688A.187S}, an updated version of the \citet{Piffaretti2011} meta-catalogue, and the first eROSITA all-sky cluster catalogue \citep{2024A&A...685A.106B}. Cross-matching is performed with deliberately strict criteria, an angular separation $\leq 1.0^\circ$ and redshift difference $\Delta z \leq 0.005$, retaining only the closest halo mass match when multiple candidates qualify (this ambiguity arises only 4 and 1 times for the MCXC and eROSITA catalogues respectively, less than 5\%). This prioritizes secure, high-quality identifications.

Before examining mass comparisons, we quantify the catalogue's recovery of known X-ray clusters. Applying the cross-matching criteria within the volume $z \leq 0.05$ and $|b| > 10^\circ$, \manticorelocal recovers 70 of 313 MCXC-II clusters (22\%) and 33 of 165 eROSITA clusters (20\%). These modest overall fractions reflect a fundamental mass threshold mismatch: \manticorelocal targets systems with $\langle M_{200} \rangle > 10^{14}$~\msol, while the X-ray catalogues extend to substantially lower masses. Recovery is strongly mass-dependent: at $M_{500} > 2 \times 10^{14}$~\msol, \manticorelocal recovers 47\% of MCXC-II and 40\% of eROSITA clusters, rising to 100\% for eROSITA systems above $5 \times 10^{14}$~\msol.

Conversely, 73\% of \manticorelocal associations lack X-ray counterparts in either catalogue. However, these unmatched systems are predominantly low-mass ($M_{500} < 10^{14}$~\msol) and at larger distances ($R > 150$~Mpc), consistent with falling below X-ray detection thresholds rather than being spurious. At $\langle M_{500} \rangle > 2 \times 10^{14}$~\msol, only 6\% of associations lack X-ray matches, and all six associations above $5 \times 10^{14}$~\msol\ have confirmed X-ray counterparts. This mass-dependent purity suggests that the unmatched low-mass associations represent real structures below X-ray sensitivity limits.

Positional offsets between matched \manticorelocal associations and their X-ray counterparts are consistent with the posterior uncertainties. The median angular separation is $0.33^\circ$ for MCXC-II and $0.35^\circ$ for eROSITA matches. When normalized by the posterior angular uncertainty $\sigma_\theta$ (derived from the 3D positional scatter), the median offset is $\approx 0.4\sigma$ for both catalogues, with nearly all matches falling within $1\sigma$. That the observed offsets are systematically smaller than the quoted uncertainties indicates that the posterior positional uncertainties are correctly estimated and, if anything, slightly conservative.

\Cref{fig:mass_ratio_external} shows the ratio of \textsc{Manticore-Local} $M_{500}$ masses to the corresponding external values. Error bars on individual points represent the 25th--75th percentile range of the \manticorelocal posterior mass distribution; we do not propagate measurement uncertainties from the external catalogues, which are typically 30--50\% for MCXC-II \citep[combining hydrostatic bias and scaling relation scatter;][]{Piffaretti2011} and 25--35\% for eROSITA \citep{2024A&A...685A.106B}. By comparison, the median fractional uncertainty on \manticorelocal masses is $\approx 25\%$ (0.11~dex), indicating that the posterior provides constraints comparable to or tighter than typical X-ray estimates. Both the fiducial and strict catalogue selections yield consistent mass ratios, demonstrating that the results are robust to the choice of selection criteria.

We find that \textsc{Manticore-Local} masses are systematically higher than MCXC-II, with a median ratio of $\sim 1.7$. The ratio distribution is broad and positively skewed, with individual outliers reaching ratios of 3--9 (visible as scattered points above the shaded interquartile range in \cref{fig:mass_ratio_external}). In contrast, the comparison to eROSITA yields a median ratio close to unity with only a mild positive bias and no strong mass dependence.

To contextualize these differences, we note that MCXC-II and eROSITA masses themselves disagree substantially for clusters common to both catalogues: within our matching volume, the median MCXC/eROSITA mass ratio is $\approx 0.54$, with 93\% of common clusters having lower MCXC masses. This baseline disagreement between X-ray catalogues---consistent with the expected hydrostatic bias in MCXC that eROSITA's weak-lensing calibration is designed to correct---provides essential context for interpreting the \manticorelocal comparisons.

The larger offsets with MCXC-II exceed the 10--30\% hydrostatic bias typically quoted for individual clusters \citep[e.g.][]{2006MNRAS.369.2013R,2007ApJ...655...98N}. However, several compounding factors contribute to this discrepancy. First, MCXC-II masses are not direct hydrostatic measurements but are derived from $L_X$--$M$ and $Y_X$--$M$ scaling relations calibrated using small samples of predominantly relaxed clusters \citep{Piffaretti2011,2024A&A...688A.187S}. This introduces additional systematic uncertainties beyond the hydrostatic bias itself, as the scaling relations may not accurately represent the full cluster population, particularly dynamically disturbed systems. Second, mock X-ray analyses of modern cosmological simulations demonstrate that the \textit{combined} effect of hydrostatic assumptions, spectroscopic systematics, and scaling relation scatter can produce biases exceeding 30\% for massive clusters \citep{2021MNRAS.506.2533B,2024MNRAS.534..251K}, with the most disturbed systems showing even larger offsets. Third, cosmological analyses attempting to reconcile X-ray cluster counts with CMB and SZ measurements have inferred effective biases as large as $(1-b) \sim 0.6$, corresponding to a factor of $\sim 1.7$ underestimation \citep[e.g.][]{2016A&A...594A..24P,Medezinski2018}. The broad, positively skewed distribution of mass ratios we observe (\cref{fig:mass_ratio_external}) suggests that multiple systematic effects contribute at varying levels across the cluster population, with the largest discrepancies occurring for the most massive and potentially most disturbed systems where all biases are amplified.

By contrast, eROSITA explicitly calibrates its mass proxy using weak-lensing measurements from DES, KiDS, and HSC \citep{2024A&A...689A.298G}, anchoring the X-ray mass scale directly to gravitational lensing. This calibration is designed to bring eROSITA masses closer to the total gravitational mass, a quantity that \textsc{Manticore-Local} also aims to measure through forward-modelled gravitational dynamics in the N-body simulations. The near-unity median ratio between \textsc{Manticore-Local} and eROSITA is consistent with both methods targeting the same underlying mass definition, though we note that each approach carries its own systematic uncertainties.

A subset of systems remain strong outliers (mass ratio $>3$ or $<0.33$) in both comparisons: 14\% of MCXC-II matches and 13\% of eROSITA matches. The origin of these discrepancies---whether from X-ray measurement systematics, reconstruction limitations, or genuine physical effects---cannot be determined from mass ratios alone. To investigate, we examined whether the same objects appear as outliers in both catalogues: among the 23 clusters matched in both MCXC-II and eROSITA, only 2 are outliers in both comparisons, while 4 are outliers in only one catalogue. This lack of overlap suggests that outliers are primarily driven by catalogue-specific systematics rather than failures in the \manticorelocal reconstruction.

We further tested whether outliers show signs of reduced reconstruction quality by comparing their posterior properties to well-matched systems. Outliers do not have statistically significantly broader mass posteriors (Mann-Whitney $p > 0.3$) or larger positional scatter ($p > 0.06$). If anything, outliers tend to have slightly \textit{narrower} posteriors and higher membership counts than well-matched clusters, indicating that they are not poorly constrained associations. While not definitive, this pattern is more consistent with limitations in X-ray mass proxies for these specific systems, potentially due to complex dynamical states or projection effects, than with reconstruction failures in \manticorelocal.

Nonetheless, we cannot definitively attribute these outliers to observational systematics without independent validation. Constrained catalogues provide a framework to flag such systems for targeted follow-up, where detailed multi-wavelength observations can diagnose the origin of discrepancies. This capacity to identify individual objects with significant mass discrepancies, rather than marginalizing over them in population-level fits, is an advantage of posterior-based reconstructions, enabling systematic investigation of potential biases in cluster mass proxies.

\section{Conclusions}
\label{sect:conclusions}

We have presented a posterior catalogue of cluster-scale halo associations constructed from the \textsc{Manticore-Local} inference of the nearby Universe. This catalogue builds directly on the validation framework of \citet{McAlpine2025}, extending the analysis from statistical characterizations of the reconstructed density field to a catalogue of individually identified systems. It links present-day observational constraints to the inferred positions, masses, and evolutionary histories of massive structures, providing the community with a validated set of digital analogues to real clusters in the local Universe. The full catalogue, including posterior distributions of key properties for all identified systems, is publicly available at \href{https://cosmictwin.org}{\texttt{cosmictwin.org}}.

The principal outcomes of this study are:

\begin{itemize}

\item \textbf{Identification of posterior associations:}  
We have defined and catalogued ``halo associations'' as coherent structures that remain clustered across the posterior ensemble, representing real physical systems inferred from present-day galaxy observations. The fiducial catalogue contains 401 associations with $\langle M_{200} \rangle \geq 10^{14}\,{\rm M_\odot}$, selected by requiring ambiguity rates below 5\% and at least 20 member haloes; a strict sample of 162 associations with additional constraints on positional and mass scatter is also provided. Each association includes posterior distributions of position, mass, and velocity. This establishes a bridge between field-level inference and traditional cluster catalogues, delivering a physically grounded, uncertainty-aware sample of massive structures.

\item \textbf{Validation via independent observables:}
Stacked \textit{Planck} PR4 NILC Compton-$y$ measurements yield statistically significant thermal Sunyaev--Zel'dovich detections at the \textsc{Manticore-Local} association positions. This constitutes a posterior predictive test: the tSZ field plays no role in the inference, so the presence of a coherent decrement at the predicted halo locations provides evidence that the associations trace genuine massive structures on average.

Across increasing mass bins, the stacked \textsc{Manticore-Local} signals show the expected mass trend, consistent with the established $M_{500}$--$Y$ scaling relation, producing larger decrements for higher-mass systems. The absolute amplitudes are modestly suppressed, which is naturally explained by the $\sim$Mpc-level positional scatter in the posterior ensemble that smooths and broadens the stacked profiles. Crucially, the detections arise solely from the reconstructed halo positions and masses, without any tuning to match external catalogues or observed SZ amplitudes. The tSZ measurements therefore provide ensemble-level validation that the reconstructed associations trace hot intracluster gas rather than being artefacts of the optical inference; they do not, however, validate individual associations or test positional accuracy at the object level.

\item \textbf{Constraints on evolutionary pathways:}  
By tracing halo merger trees back to $z=5$, we find that association progenitors occupy compact and anisotropic regions of Lagrangian space, exhibiting coherent infall along preferred directions. Comparisons with mass- and displacement-matched control haloes from unconstrained simulations show that the inference reduces allowed configuration-space volumes by typical factors of $2$--$5$, demonstrating that present-day galaxy data alone, even without direct velocity measurements or high-redshift observations, constrain not only final cluster properties but also aspects of their formation histories.

These findings naturally connect to the recent analysis of \citet{2024MNRAS.534.3120S} on the Lagrangian overlap of constrained simulations, which quantified, at the population level, how observational constraints restrict the accessible regions of initial condition space. While that work established the effect in terms of global overlap statistics, the present study demonstrates the same principle at the level of individual objects, showing directly that the evolutionary pathways of specific cluster associations are localized and non-random. Together, these complementary approaches reinforce the conclusion that field-level inference encodes genuine information about cosmic history: constrained reconstructions do not merely reproduce present-day structures, but systematically narrow the range of possible past configurations that could have led to them.

\item \textbf{External mass scale comparisons:}
One-to-one cross-matches with X-ray catalogues recover known systematic offsets. Relative to MCXC-II, \textsc{Manticore-Local} masses are typically higher by factors of $\sim 1.7$, consistent with the compounding effects of hydrostatic bias and scaling relation systematics in X-ray mass proxies, with individual outliers reaching ratios of 3--9. By contrast, agreement with the weak-lensing-calibrated eROSITA mass scale is close to unity. Outlier diagnostics suggest that mass discrepancies are more likely driven by catalogue-specific X-ray systematics than by reconstruction failures, though definitive attribution requires independent validation. Constrained catalogues thus provide a framework to flag individual systems with significant discrepancies for targeted follow-up, enabling systematic investigation of potential biases rather than marginalizing over them in population-level analyses.

\end{itemize}

The \textsc{Manticore-Local} cluster catalogue thus provides a robust and observationally validated representation of massive structures in the nearby Universe. For reference, we summarize its key quality metrics: completeness is strongly mass- and distance-dependent, with recovery fractions increasing from $\sim$20\% for all X-ray clusters to $>$90\% for the most massive systems ($M_{500} > 5 \times 10^{14}$~\msol), while degrading beyond $\sim$150~Mpc as the underlying \tmpp\ constraints weaken. Positional uncertainties are well-calibrated, with median angular offsets of $\sim 0.3^\circ$ corresponding to $\approx 0.4\sigma$ of the posterior uncertainty. Mass uncertainties are $\approx$25\% (0.11~dex), comparable to or tighter than typical X-ray estimates. These metrics should guide users in assessing the reliability of individual associations as a function of mass and distance.

The catalogue's key strength lies in providing digital counterparts to real observed clusters, each with validated positions, masses, and formation histories. Because it is rooted in the posterior ensemble, the catalogue encodes observational uncertainties in a transparent way, allowing them to be propagated rigorously into downstream analyses.

An important capability enabled by this framework is direct, object-by-object comparison between simulated and observed clusters. Unlike traditional population-level simulation studies that rely on ensemble-averaged scaling relations, constrained catalogues allow mismatches to be localized to specific systems, where astrophysical or methodological origins can potentially be investigated. This shifts the paradigm from identifying general trends to flagging particular discrepancies: outliers become candidates for targeted investigation rather than statistical noise to be marginalized over. While definitive attribution of discrepancies to observational systematics versus reconstruction limitations requires independent validation, this framework provides a starting point for such investigations. Hydrodynamical follow-up simulations of the posterior initial conditions will extend this capability to comparisons of X-ray emission, SZ profiles, and galaxy properties.

The catalogue also enables hybrid analyses that combine observed and simulated quantities in complementary ways. As demonstrated in our tSZ validation, one can use observed sky positions for stacking analyses while drawing masses, velocities, and formation histories from the constrained simulation. This approach can be extended to probes such as the kinetic SZ and moving-lens effects, which require accurate velocity information that is inaccessible observationally but naturally available in constrained reconstructions.

The present catalogue forms part of a growing \textsc{Manticore} product suite. Alongside the local void catalogue introduced in \citet{2025arXiv250706866M}, it delivers complementary perspectives on the most underdense and overdense environments of the nearby Universe. Forthcoming expansions, particularly \textsc{Manticore-Deep}, will extend the framework to constrained volumes approximately two orders of magnitude larger than \textsc{Manticore-Local}, enabling population-level studies of cluster properties, environmental dependencies, and rare massive systems that are underrepresented in the local volume.

Looking ahead, this programme represents a step toward comprehensive digital twins of the local Universe, where galaxies, clusters, and voids are reconstructed within a physically self-consistent posterior framework. Such digital twins will enable the community to test models of galaxy and cluster formation with unprecedented realism, perform joint analyses that fully integrate observational and simulated data, and develop new diagnostic tools that exploit the unique combination of observational anchoring and forward-modelled physics. The \textsc{Manticore-Local} cluster catalogue, validated through independent observables and released as a community resource, provides a foundation for this next generation of cosmological analysis.

\section*{Acknowledgements}

SM thanks Ludvig Doeser and Richard Stiskalek for their useful inputs and discussions. SM also thanks the anonymous referee for their careful reading and constructive comments, which significantly improved the quality and clarity of this manuscript.

We acknowledge computational resources provided by the Swedish National Infrastructure for Computing (SNIC) at the PDC Center for High Performance Computing, KTH Royal Institute of Technology, partially funded by the Swedish Research Council through grant agreement no. 2018-05973. This work used the DiRAC@Durham facility managed by the Institute for Computational Cosmology on behalf of the STFC DiRAC HPC Facility (www.dirac.ac.uk). The equipment was funded by BEIS capital funding via STFC capital grants ST/K00042X/1, ST/P002293/1, ST/R002371/1 and ST/S002502/1, Durham University and STFC operations grant ST/R000832/1. DiRAC is part of the National e-Infrastructure. In addition, this work has made use of the Infinity Cluster hosted by Institut d'Astrophysique de Paris, and was granted access to the HPC resources of TGCC (Très Grand Centre de Calcul), Irene-Joliot-Curie supercomputer, under the allocations A0170415682 and SS010415380.

SM acknowledges support from the Simons Foundation through the Simons Collaboration on "Learning the Universe".

This work is done within the Aquila Consortium\footnote{\url{https://www.aquila-consortium.org/}}, Virgo Consortium\footnote{\url{https://virgo.dur.ac.uk/}} and Learning the Universe Collaboration\footnote{\url{https://learning-the-universe.org/}}.

%%%%%%%%%%%%%%%%%%%%%%%%%%%%%%%%%%%%%%%%%%%%%%%%%%
\section*{Data Availability}

The posterior association catalogue presented in this work is publicly available at \href{https://cosmictwin.org}{\texttt{cosmictwin.org}}. The catalogue includes positions, masses, velocities, member halo lists, and summary statistics for all 401 fiducial associations, along with full posterior distributions for key properties. Data products are provided in machine-readable format with comprehensive documentation describing the data structure and usage guidelines, enabling direct integration into external analyses and cross-correlation studies.

The code used to construct the catalogue and generate the figures in this paper is available \href{https://github.com/stuartmcalpine/manticore_posterior_clusters}{here}.

A demonstration Jupyter notebook showing how to load, query, and visualize the posterior association catalogue is available \href{https://github.com/stuartmcalpine/manticore_public_scripts}{here}, providing worked examples for common use cases.

%%%%%%%%%%%%%%%%%%%% REFERENCES %%%%%%%%%%%%%%%%%%

\bibliographystyle{mnras}
\bibliography{example} % if your bibtex file is called example.bib

@ARTICLE{2010MNRAS.408.2163A,
       author = {{Arag{\'o}n-Calvo}, Miguel A. and {van de Weygaert}, Rien and {Jones}, Bernard J.~T.},
        title = "{Multiscale phenomenology of the cosmic web}",
      journal = {\mnras},
         year = 2010,
        month = nov,
       volume = {408},
       number = {4},
        pages = {2163-2187},
          doi = {10.1111/j.1365-2966.2010.17263.x},
archivePrefix = {arXiv},
       eprint = {1007.0742},
 primaryClass = {astro-ph.CO},
       adsurl = {https://ui.adsabs.harvard.edu/abs/2010MNRAS.408.2163A}
}

@ARTICLE{Wang2014,
       author = {{Wang}, Huiyuan and {Mo}, H.~J. and {Yang}, Xiaohu and {Jing}, Y.~P. and {Lin}, W.~P.},
        title = "{ELUCID{\textemdash}Exploring the Local Universe with the Reconstructed Initial Density Field. I. Hamiltonian Markov Chain Monte Carlo Method with Particle Mesh Dynamics}",
      journal = {\apj},
         year = 2014,
        month = oct,
       volume = {794},
       number = {1},
          eid = {94},
        pages = {94},
          doi = {10.1088/0004-637X/794/1/94},
archivePrefix = {arXiv},
       eprint = {1407.3451},
 primaryClass = {astro-ph.CO},
       adsurl = {https://ui.adsabs.harvard.edu/abs/2014ApJ...794...94W}
}

@ARTICLE{Wang2016,
       author = {{Wang}, Huiyuan and {Mo}, H.~J. and {Yang}, Xiaohu and {Zhang}, Youcai and {Shi}, JingJing and {Jing}, Y.~P. and {Liu}, Chengze and {Li}, Shijie and {Kang}, Xi and {Gao}, Yang},
        title = "{ELUCID - Exploring the Local Universe with ReConstructed Initial Density Field III: Constrained Simulation in the SDSS Volume}",
      journal = {\apj},
         year = 2016,
        month = nov,
       volume = {831},
       number = {2},
          eid = {164},
        pages = {164},
          doi = {10.3847/0004-637X/831/2/164},
archivePrefix = {arXiv},
       eprint = {1608.01763},
 primaryClass = {astro-ph.CO},
       adsurl = {https://ui.adsabs.harvard.edu/abs/2016ApJ...831..164W}
}

@ARTICLE{Ferrarese2012,
       author = {{Ferrarese}, Laura and {C{\^o}t{\'e}}, Patrick and {Cuillandre}, Jean-Charles and {Gwyn}, S.~D.~J. and {Peng}, Eric W. and {MacArthur}, Lauren A. and {Duc}, Pierre-Alain and {Boselli}, A. and {Mei}, Simona and {Erben}, Thomas and {McConnachie}, Alan W. and {Durrell}, Patrick R. and {Mihos}, J. Christopher and {Jord{\'a}n}, Andr{\'e}s and {Lan{\c{c}}on}, Ariane and {Puzia}, Thomas H. and {Emsellem}, Eric and {Balogh}, Michael L. and {Blakeslee}, John P. and {van Waerbeke}, Ludovic and {Gavazzi}, Rapha{\"e}l and {Vollmer}, Bernd and {Kavelaars}, J.~J. and {Woods}, David and {Ball}, Nicholas M. and {Boissier}, S. and {Courteau}, St{\'e}phane and {Ferriere}, E. and {Gavazzi}, G. and {Hildebrandt}, Hendrik and {Hudelot}, P. and {Huertas-Company}, M. and {Liu}, Chengze and {McLaughlin}, Dean and {Mellier}, Y. and {Milkeraitis}, Martha and {Schade}, David and {Balkowski}, Chantal and {Bournaud}, Fr{\'e}d{\'e}ric and {Carlberg}, R.~G. and {Chapman}, S.~C. and {Hoekstra}, Henk and {Peng}, Chien and {Sawicki}, Marcin and {Simard}, Luc and {Taylor}, James E. and {Tully}, R. Brent and {van Driel}, Wim and {Wilson}, Christine D. and {Burdullis}, Todd and {Mahoney}, Billy and {Manset}, Nadine},
        title = "{The Next Generation Virgo Cluster Survey (NGVS). I. Introduction to the Survey}",
      journal = {\apjs},
         year = 2012,
        month = may,
       volume = {200},
       number = {1},
          eid = {4},
        pages = {4},
          doi = {10.1088/0067-0049/200/1/4},
       adsurl = {https://ui.adsabs.harvard.edu/abs/2012ApJS..200....4F}
}

@ARTICLE{Planck2016,
       author = {{Planck Collaboration} and {Ade}, P.~A.~R. and {Aghanim}, N. and {Arnaud}, M. and {Ashdown}, M. and {Aumont}, J. and {Baccigalupi}, C. and {Banday}, A.~J. and {Barreiro}, R.~B. and {Bartlett}, J.~G. and {Bartolo}, N. and {Battaner}, E. and {Battye}, R. and {Benabed}, K. and {Beno{\^\i}t}, A. and {Benoit-L{\'e}vy}, A. and {Bernard}, J. -P. and {Bersanelli}, M. and {Bielewicz}, P. and {Bock}, J.~J. and {Bonaldi}, A. and {Bonavera}, L. and {Bond}, J.~R. and {Borrill}, J. and {Bouchet}, F.~R. and {Boulanger}, F. and {Bucher}, M. and {Burigana}, C. and {Butler}, R.~C. and {Calabrese}, E. and {Cardoso}, J. -F. and {Catalano}, A. and {Challinor}, A. and {Chamballu}, A. and {Chary}, R. -R. and {Chiang}, H.~C. and {Chluba}, J. and {Christensen}, P.~R. and {Church}, S. and {Clements}, D.~L. and {Colombi}, S. and {Colombo}, L.~P.~L. and {Combet}, C. and {Coulais}, A. and {Crill}, B.~P. and {Curto}, A. and {Cuttaia}, F. and {Danese}, L. and {Davies}, R.~D. and {Davis}, R.~J. and {de Bernardis}, P. and {de Rosa}, A. and {de Zotti}, G. and {Delabrouille}, J. and {D{\'e}sert}, F. -X. and {Di Valentino}, E. and {Dickinson}, C. and {Diego}, J.~M. and {Dolag}, K. and {Dole}, H. and {Donzelli}, S. and {Dor{\'e}}, O. and {Douspis}, M. and {Ducout}, A. and {Dunkley}, J. and {Dupac}, X. and {Efstathiou}, G. and {Elsner}, F. and {En{\ss}lin}, T.~A. and {Eriksen}, H.~K. and {Farhang}, M. and {Fergusson}, J. and {Finelli}, F. and {Forni}, O. and {Frailis}, M. and {Fraisse}, A.~A. and {Franceschi}, E. and {Frejsel}, A. and {Galeotta}, S. and {Galli}, S. and {Ganga}, K. and {Gauthier}, C. and {Gerbino}, M. and {Ghosh}, T. and {Giard}, M. and {Giraud-H{\'e}raud}, Y. and {Giusarma}, E. and {Gjerl{\o}w}, E. and {Gonz{\'a}lez-Nuevo}, J. and {G{\'o}rski}, K.~M. and {Gratton}, S. and {Gregorio}, A. and {Gruppuso}, A. and {Gudmundsson}, J.~E. and {Hamann}, J. and {Hansen}, F.~K. and {Hanson}, D. and {Harrison}, D.~L. and {Helou}, G. and {Henrot-Versill{\'e}}, S. and {Hern{\'a}ndez-Monteagudo}, C. and {Herranz}, D. and {Hildebrandt}, S.~R. and {Hivon}, E. and {Hobson}, M. and {Holmes}, W.~A. and {Hornstrup}, A. and {Hovest}, W. and {Huang}, Z. and {Huffenberger}, K.~M. and {Hurier}, G. and {Jaffe}, A.~H. and {Jaffe}, T.~R. and {Jones}, W.~C. and {Juvela}, M. and {Keih{\"a}nen}, E. and {Keskitalo}, R. and {Kisner}, T.~S. and {Kneissl}, R. and {Knoche}, J. and {Knox}, L. and {Kunz}, M. and {Kurki-Suonio}, H. and {Lagache}, G. and {L{\"a}hteenm{\"a}ki}, A. and {Lamarre}, J. -M. and {Lasenby}, A. and {Lattanzi}, M. and {Lawrence}, C.~R. and {Leahy}, J.~P. and {Leonardi}, R. and {Lesgourgues}, J. and {Levrier}, F. and {Lewis}, A. and {Liguori}, M. and {Lilje}, P.~B. and {Linden-V{\o}rnle}, M. and {L{\'o}pez-Caniego}, M. and {Lubin}, P.~M. and {Mac{\'\i}as-P{\'e}rez}, J.~F. and {Maggio}, G. and {Maino}, D. and {Mandolesi}, N. and {Mangilli}, A. and {Marchini}, A. and {Maris}, M. and {Martin}, P.~G. and {Martinelli}, M. and {Mart{\'\i}nez-Gonz{\'a}lez}, E. and {Masi}, S. and {Matarrese}, S. and {McGehee}, P. and {Meinhold}, P.~R. and {Melchiorri}, A. and {Melin}, J. -B. and {Mendes}, L. and {Mennella}, A. and {Migliaccio}, M. and {Millea}, M. and {Mitra}, S. and {Miville-Desch{\^e}nes}, M. -A. and {Moneti}, A. and {Montier}, L. and {Morgante}, G. and {Mortlock}, D. and {Moss}, A. and {Munshi}, D. and {Murphy}, J.~A. and {Naselsky}, P. and {Nati}, F. and {Natoli}, P. and {Netterfield}, C.~B. and {N{\o}rgaard-Nielsen}, H.~U. and {Noviello}, F. and {Novikov}, D. and {Novikov}, I. and {Oxborrow}, C.~A. and {Paci}, F. and {Pagano}, L. and {Pajot}, F. and {Paladini}, R. and {Paoletti}, D. and {Partridge}, B. and {Pasian}, F. and {Patanchon}, G. and {Pearson}, T.~J. and {Perdereau}, O. and {Perotto}, L. and {Perrotta}, F. and {Pettorino}, V. and {Piacentini}, F. and {Piat}, M. and {Pierpaoli}, E. and {Pietrobon}, D. and {Plaszczynski}, S. and {Pointecouteau}, E. and {Polenta}, G. and {Popa}, L. and {Pratt}, G.~W. and {Pr{\'e}zeau}, G. and {Prunet}, S. and {Puget}, J. -L. and {Rachen}, J.~P. and {Reach}, W.~T. and {Rebolo}, R. and {Reinecke}, M. and {Remazeilles}, M. and {Renault}, C. and {Renzi}, A. and {Ristorcelli}, I. and {Rocha}, G. and {Rosset}, C. and {Rossetti}, M. and {Roudier}, G. and {Rouill{\'e} d'Orfeuil}, B. and {Rowan-Robinson}, M. and {Rubi{\~n}o-Mart{\'\i}n}, J.~A. and {Rusholme}, B. and {Said}, N. and {Salvatelli}, V. and {Salvati}, L. and {Sandri}, M. and {Santos}, D. and {Savelainen}, M. and {Savini}, G. and {Scott}, D. and {Seiffert}, M.~D. and {Serra}, P. and {Shellard}, E.~P.~S. and {Spencer}, L.~D. and {Spinelli}, M. and {Stolyarov}, V. and {Stompor}, R. and {Sudiwala}, R. and {Sunyaev}, R. and {Sutton}, D. and {Suur-Uski}, A. -S. and {Sygnet}, J. -F. and {Tauber}, J.~A. and {Terenzi}, L. and {Toffolatti}, L. and {Tomasi}, M. and {Tristram}, M. and {Trombetti}, T. and {Tucci}, M. and {Tuovinen}, J. and {T{\"u}rler}, M. and {Umana}, G. and {Valenziano}, L. and {Valiviita}, J. and {Van Tent}, F. and {Vielva}, P. and {Villa}, F. and {Wade}, L.~A. and {Wandelt}, B.~D. and {Wehus}, I.~K. and {White}, M. and {White}, S.~D.~M. and {Wilkinson}, A. and {Yvon}, D. and {Zacchei}, A. and {Zonca}, A.},
        title = "{Planck 2015 results. XIII. Cosmological parameters}",
      journal = {\aap},
         year = 2016,
        month = sep,
       volume = {594},
          eid = {A13},
        pages = {A13},
          doi = {10.1051/0004-6361/201525830},
archivePrefix = {arXiv},
       eprint = {1502.01589},
 primaryClass = {astro-ph.CO},
       adsurl = {https://ui.adsabs.harvard.edu/abs/2016A&A...594A..13P}
}

@ARTICLE{Rines2003,
       author = {{Rines}, Kenneth and {Geller}, Margaret J. and {Kurtz}, Michael J. and {Diaferio}, Antonaldo},
        title = "{CAIRNS: The Cluster and Infall Region Nearby Survey. I. Redshifts and Mass Profiles}",
      journal = {\aj},
         year = 2003,
        month = nov,
       volume = {126},
       number = {5},
        pages = {2152-2170},
          doi = {10.1086/378599},
archivePrefix = {arXiv},
       eprint = {astro-ph/0306538},
 primaryClass = {astro-ph},
       adsurl = {https://ui.adsabs.harvard.edu/abs/2003AJ....126.2152R}
}

@ARTICLE{Piffaretti2011,
       author = {{Piffaretti}, R. and {Arnaud}, M. and {Pratt}, G.~W. and {Pointecouteau}, E. and {Melin}, J. -B.},
        title = "{The MCXC: a meta-catalogue of x-ray detected clusters of galaxies}",
      journal = {\aap},
         year = 2011,
        month = oct,
       volume = {534},
          eid = {A109},
        pages = {A109},
          doi = {10.1051/0004-6361/201015377},
archivePrefix = {arXiv},
       eprint = {1007.1916},
 primaryClass = {astro-ph.CO},
       adsurl = {https://ui.adsabs.harvard.edu/abs/2011A&A...534A.109P}
}

@ARTICLE{Jasche2013,
       author = {{Jasche}, Jens and {Wandelt}, Benjamin D.},
        title = "{Bayesian physical reconstruction of initial conditions from large-scale structure surveys}",
      journal = {\mnras},
         year = 2013,
        month = jun,
       volume = {432},
       number = {2},
        pages = {894-913},
          doi = {10.1093/mnras/stt449},
archivePrefix = {arXiv},
       eprint = {1203.3639},
 primaryClass = {astro-ph.CO},
       adsurl = {https://ui.adsabs.harvard.edu/abs/2013MNRAS.432..894J}
}

@ARTICLE{Jasche2019,
       author = {{Jasche}, J. and {Lavaux}, G.},
        title = "{Physical Bayesian modelling of the non-linear matter distribution: New insights into the nearby universe}",
      journal = {\aap},
         year = 2019,
        month = may,
       volume = {625},
          eid = {A64},
        pages = {A64},
          doi = {10.1051/0004-6361/201833710},
archivePrefix = {arXiv},
       eprint = {1806.11117},
 primaryClass = {astro-ph.CO},
       adsurl = {https://ui.adsabs.harvard.edu/abs/2019A&A...625A..64J}
}

@ARTICLE{2006Sci...313..311B,
       author = {{Bland-Hawthorn}, J. and {Peebles}, P.~J.~E.},
        title = "{Near-Field Cosmology}",
      journal = {Science},
         year = 2006,
        month = jul,
       volume = {313},
        pages = {311-312},
          doi = {10.1126/science.1127183},
       adsurl = {https://ui.adsabs.harvard.edu/abs/2006Sci...313..311B}
}

@ARTICLE{Lavaux2011,
       author = {{Lavaux}, Guilhem and {Hudson}, Michael J.},
        title = "{The 2M++ galaxy redshift catalogue}",
      journal = {\mnras},
         year = 2011,
        month = oct,
       volume = {416},
       number = {4},
        pages = {2840-2856},
          doi = {10.1111/j.1365-2966.2011.19233.x},
archivePrefix = {arXiv},
       eprint = {1105.6107},
 primaryClass = {astro-ph.CO},
       adsurl = {https://ui.adsabs.harvard.edu/abs/2011MNRAS.416.2840L}
}

@ARTICLE{Pfeifer2023,
       author = {{Pfeifer}, Simon and {Valade}, Aur{\'e}lien and {Gottl{\"o}ber}, Stefan and {Hoffman}, Yehuda and {Libeskind}, Noam I. and {Hellwing}, Wojciech A.},
        title = "{A local universe model for constrained simulations}",
      journal = {\mnras},
         year = 2023,
        month = aug,
       volume = {523},
       number = {4},
        pages = {5985-5994},
          doi = {10.1093/mnras/stad1851},
archivePrefix = {arXiv},
       eprint = {2305.05694},
 primaryClass = {astro-ph.CO},
       adsurl = {https://ui.adsabs.harvard.edu/abs/2023MNRAS.523.5985P}
}

@ARTICLE{DEScosmo,
       author = {{Abbott}, T.~M.~C. and {Aguena}, M. and {Alarcon}, A. and {Allam}, S. and {Alves}, O. and {Amon}, A. and {Andrade-Oliveira}, F. and {Annis}, J. and {Avila}, S. and {Bacon}, D. and {Baxter}, E. and {Bechtol}, K. and {Becker}, M.~R. and {Bernstein}, G.~M. and {Bhargava}, S. and {Birrer}, S. and {Blazek}, J. and {Brandao-Souza}, A. and {Bridle}, S.~L. and {Brooks}, D. and {Buckley-Geer}, E. and {Burke}, D.~L. and {Camacho}, H. and {Campos}, A. and {Carnero Rosell}, A. and {Carrasco Kind}, M. and {Carretero}, J. and {Castander}, F.~J. and {Cawthon}, R. and {Chang}, C. and {Chen}, A. and {Chen}, R. and {Choi}, A. and {Conselice}, C. and {Cordero}, J. and {Costanzi}, M. and {Crocce}, M. and {da Costa}, L.~N. and {da Silva Pereira}, M.~E. and {Davis}, C. and {Davis}, T.~M. and {De Vicente}, J. and {DeRose}, J. and {Desai}, S. and {Di Valentino}, E. and {Diehl}, H.~T. and {Dietrich}, J.~P. and {Dodelson}, S. and {Doel}, P. and {Doux}, C. and {Drlica-Wagner}, A. and {Eckert}, K. and {Eifler}, T.~F. and {Elsner}, F. and {Elvin-Poole}, J. and {Everett}, S. and {Evrard}, A.~E. and {Fang}, X. and {Farahi}, A. and {Fernandez}, E. and {Ferrero}, I. and {Fert{\'e}}, A. and {Fosalba}, P. and {Friedrich}, O. and {Frieman}, J. and {Garc{\'\i}a-Bellido}, J. and {Gatti}, M. and {Gaztanaga}, E. and {Gerdes}, D.~W. and {Giannantonio}, T. and {Giannini}, G. and {Gruen}, D. and {Gruendl}, R.~A. and {Gschwend}, J. and {Gutierrez}, G. and {Harrison}, I. and {Hartley}, W.~G. and {Herner}, K. and {Hinton}, S.~R. and {Hollowood}, D.~L. and {Honscheid}, K. and {Hoyle}, B. and {Huff}, E.~M. and {Huterer}, D. and {Jain}, B. and {James}, D.~J. and {Jarvis}, M. and {Jeffrey}, N. and {Jeltema}, T. and {Kovacs}, A. and {Krause}, E. and {Kron}, R. and {Kuehn}, K. and {Kuropatkin}, N. and {Lahav}, O. and {Leget}, P. -F. and {Lemos}, P. and {Liddle}, A.~R. and {Lidman}, C. and {Lima}, M. and {Lin}, H. and {MacCrann}, N. and {Maia}, M.~A.~G. and {Marshall}, J.~L. and {Martini}, P. and {McCullough}, J. and {Melchior}, P. and {Mena-Fern{\'a}ndez}, J. and {Menanteau}, F. and {Miquel}, R. and {Mohr}, J.~J. and {Morgan}, R. and {Muir}, J. and {Myles}, J. and {Nadathur}, S. and {Navarro-Alsina}, A. and {Nichol}, R.~C. and {Ogando}, R.~L.~C. and {Omori}, Y. and {Palmese}, A. and {Pandey}, S. and {Park}, Y. and {Paz-Chinch{\'o}n}, F. and {Petravick}, D. and {Pieres}, A. and {Plazas Malag{\'o}n}, A.~A. and {Porredon}, A. and {Prat}, J. and {Raveri}, M. and {Rodriguez-Monroy}, M. and {Rollins}, R.~P. and {Romer}, A.~K. and {Roodman}, A. and {Rosenfeld}, R. and {Ross}, A.~J. and {Rykoff}, E.~S. and {Samuroff}, S. and {S{\'a}nchez}, C. and {Sanchez}, E. and {Sanchez}, J. and {Sanchez Cid}, D. and {Scarpine}, V. and {Schubnell}, M. and {Scolnic}, D. and {Secco}, L.~F. and {Serrano}, S. and {Sevilla-Noarbe}, I. and {Sheldon}, E. and {Shin}, T. and {Smith}, M. and {Soares-Santos}, M. and {Suchyta}, E. and {Swanson}, M.~E.~C. and {Tabbutt}, M. and {Tarle}, G. and {Thomas}, D. and {To}, C. and {Troja}, A. and {Troxel}, M.~A. and {Tucker}, D.~L. and {Tutusaus}, I. and {Varga}, T.~N. and {Walker}, A.~R. and {Weaverdyck}, N. and {Wechsler}, R. and {Weller}, J. and {Yanny}, B. and {Yin}, B. and {Zhang}, Y. and {Zuntz}, J. and {DES Collaboration}},
        title = "{Dark Energy Survey Year 3 results: Cosmological constraints from galaxy clustering and weak lensing}",
      journal = {\prd},
         year = 2022,
        month = jan,
       volume = {105},
       number = {2},
          eid = {023520},
        pages = {023520},
          doi = {10.1103/PhysRevD.105.023520},
archivePrefix = {arXiv},
       eprint = {2105.13549},
 primaryClass = {astro-ph.CO},
       adsurl = {https://ui.adsabs.harvard.edu/abs/2022PhRvD.105b3520A}
}

@ARTICLE{2007MNRAS.375..489H,
       author = {{Hahn}, Oliver and {Porciani}, Cristiano and {Carollo}, C. Marcella and {Dekel}, Avishai},
        title = "{Properties of dark matter haloes in clusters, filaments, sheets and voids}",
      journal = {\mnras},
         year = 2007,
        month = feb,
       volume = {375},
       number = {2},
        pages = {489-499},
          doi = {10.1111/j.1365-2966.2006.11318.x},
archivePrefix = {arXiv},
       eprint = {astro-ph/0610280},
 primaryClass = {astro-ph},
       adsurl = {https://ui.adsabs.harvard.edu/abs/2007MNRAS.375..489H}
}

@ARTICLE{Schaller2024,
       author = {{Schaller}, Matthieu and {Borrow}, Josh and {Draper}, Peter W. and {Ivkovic}, Mladen and {McAlpine}, Stuart and {Vandenbroucke}, Bert and {Bah{\'e}}, Yannick and {Chaikin}, Evgenii and {Chalk}, Aidan B.~G. and {Chan}, Tsang Keung and {Correa}, Camila and {van Daalen}, Marcel and {Elbers}, Willem and {Gonnet}, Pedro and {Hausammann}, Lo{\"\i}c and {Helly}, John and {Hu{\v{s}}ko}, Filip and {Kegerreis}, Jacob A. and {Nobels}, Folkert S.~J. and {Ploeckinger}, Sylvia and {Revaz}, Yves and {Roper}, William J. and {Ruiz-Bonilla}, Sergio and {Sandnes}, Thomas D. and {Uyttenhove}, Yolan and {Willis}, James S. and {Xiang}, Zhen},
        title = "{SWIFT: A modern highly-parallel gravity and smoothed particle hydrodynamics solver for astrophysical and cosmological applications}",
      journal = {\mnras},
         year = 2024,
        month = may,
       volume = {530},
       number = {2},
        pages = {2378-2419},
          doi = {10.1093/mnras/stae922},
archivePrefix = {arXiv},
       eprint = {2305.13380},
 primaryClass = {astro-ph.IM},
       adsurl = {https://ui.adsabs.harvard.edu/abs/2024MNRAS.530.2378S}
}

@ARTICLE{Gottloeber2010,
       author = {{Gottloeber}, Stefan and {Hoffman}, Yehuda and {Yepes}, Gustavo},
        title = "{Constrained Local UniversE Simulations (CLUES)}",
      journal = {arXiv e-prints},
         year = 2010,
        month = may,
          eid = {arXiv:1005.2687},
        pages = {arXiv:1005.2687},
          doi = {10.48550/arXiv.1005.2687},
archivePrefix = {arXiv},
       eprint = {1005.2687},
 primaryClass = {astro-ph.CO},
       adsurl = {https://ui.adsabs.harvard.edu/abs/2010arXiv1005.2687G}
}

@ARTICLE{Stopyra2024_COLA,
       author = {{Stopyra}, Stephen and {Peiris}, Hiranya V. and {Pontzen}, Andrew and {Jasche}, Jens and {Lavaux}, Guilhem},
        title = "{Towards accurate field-level inference of massive cosmic structures}",
      journal = {\mnras},
         year = 2024,
        month = jan,
       volume = {527},
       number = {1},
        pages = {1244-1256},
          doi = {10.1093/mnras/stad3170},
archivePrefix = {arXiv},
       eprint = {2304.09193},
 primaryClass = {astro-ph.CO},
       adsurl = {https://ui.adsabs.harvard.edu/abs/2024MNRAS.527.1244S}
}

@ARTICLE{Stiskalek2025,
       author = {{Stiskalek}, Richard and {Desmond}, Harry and {Devriendt}, Julien and {Slyz}, Adrianne and {Lavaux}, Guilhem and {Hudson}, Michael J. and {Bartlett}, Deaglan J. and {Courtois}, H{\'e}l{\`e}ne M.},
        title = "{The Velocity Field Olympics: Assessing velocity field reconstructions with direct distance tracers}",
      journal = {arXiv e-prints},
         year = 2025,
        month = jan,
          eid = {arXiv:2502.00121},
        pages = {arXiv:2502.00121},
          doi = {10.48550/arXiv.2502.00121},
archivePrefix = {arXiv},
       eprint = {2502.00121},
 primaryClass = {astro-ph.CO},
       adsurl = {https://ui.adsabs.harvard.edu/abs/2025arXiv250200121S}
}

@ARTICLE{Dolag2023,
       author = {{Dolag}, Klaus and {Sorce}, Jenny G. and {Pilipenko}, Sergey and {Hern{\'a}ndez-Mart{\'\i}nez}, Elena and {Valentini}, Milena and {Gottl{\"o}ber}, Stefan and {Aghanim}, Nabila and {Khabibullin}, Ildar},
        title = "{Simulating the LOcal Web (SLOW). I. Anomalies in the local density field}",
      journal = {\aap},
         year = 2023,
        month = sep,
       volume = {677},
          eid = {A169},
        pages = {A169},
          doi = {10.1051/0004-6361/202346213},
archivePrefix = {arXiv},
       eprint = {2302.10960},
 primaryClass = {astro-ph.CO},
       adsurl = {https://ui.adsabs.harvard.edu/abs/2023A&A...677A.169D}
}

@ARTICLE{Press1974,
       author = {{Press}, William H. and {Schechter}, Paul},
        title = "{Formation of Galaxies and Clusters of Galaxies by Self-Similar Gravitational Condensation}",
      journal = {\apj},
         year = 1974,
        month = feb,
       volume = {187},
        pages = {425-438},
          doi = {10.1086/152650},
       adsurl = {https://ui.adsabs.harvard.edu/abs/1974ApJ...187..425P}
}

@ARTICLE{Medezinski2018,
       author = {{Medezinski}, Elinor and {Battaglia}, Nicholas and {Umetsu}, Keiichi and {Oguri}, Masamune and {Miyatake}, Hironao and {Nishizawa}, Atsushi J. and {Sif{\'o}n}, Crist{\'o}bal and {Spergel}, David N. and {Chiu}, I. -Non and {Lin}, Yen-Ting and {Bahcall}, Neta and {Komiyama}, Yutaka},
        title = "{Planck Sunyaev-Zel'dovich cluster mass calibration using Hyper Suprime-Cam weak lensing}",
      journal = {\pasj},
         year = 2018,
        month = jan,
       volume = {70},
          eid = {S28},
        pages = {S28},
          doi = {10.1093/pasj/psx128},
archivePrefix = {arXiv},
       eprint = {1706.00434},
 primaryClass = {astro-ph.CO},
       adsurl = {https://ui.adsabs.harvard.edu/abs/2018PASJ...70S..28M}
}

@ARTICLE{Bond1996,
       author = {{Bond}, J. Richard and {Kofman}, Lev and {Pogosyan}, Dmitry},
        title = "{How filaments of galaxies are woven into the cosmic web}",
      journal = {\nat},
         year = 1996,
        month = apr,
       volume = {380},
       number = {6575},
        pages = {603-606},
          doi = {10.1038/380603a0},
archivePrefix = {arXiv},
       eprint = {astro-ph/9512141},
 primaryClass = {astro-ph},
       adsurl = {https://ui.adsabs.harvard.edu/abs/1996Natur.380..603B}
}

@ARTICLE{Carrick2015,
       author = {{Carrick}, Jonathan and {Turnbull}, Stephen J. and {Lavaux}, Guilhem and {Hudson}, Michael J.},
        title = "{Cosmological parameters from the comparison of peculiar velocities with predictions from the 2M++ density field}",
      journal = {\mnras},
         year = 2015,
        month = jun,
       volume = {450},
       number = {1},
        pages = {317-332},
          doi = {10.1093/mnras/stv547},
archivePrefix = {arXiv},
       eprint = {1504.04627},
 primaryClass = {astro-ph.CO},
       adsurl = {https://ui.adsabs.harvard.edu/abs/2015MNRAS.450..317C}
}

@ARTICLE{2024MNRAS.534.3120S,
       author = {{Stiskalek}, Richard and {Desmond}, Harry and {Devriendt}, Julien and {Slyz}, Adrianne},
        title = "{Evaluating the variance of individual halo properties in constrained cosmological simulations}",
      journal = {\mnras},
         year = 2024,
        month = nov,
       volume = {534},
       number = {4},
        pages = {3120-3132},
          doi = {10.1093/mnras/stae2292},
archivePrefix = {arXiv},
       eprint = {2310.20672},
 primaryClass = {astro-ph.CO},
       adsurl = {https://ui.adsabs.harvard.edu/abs/2024MNRAS.534.3120S}
}

@ARTICLE{McAlpine2025,
       author = {{McAlpine}, Stuart and {Jasche}, Jens and {Ata}, Metin and {Lavaux}, Guilhem and {Stiskalek}, Richard and {Frenk}, Carlos S. and {Jenkins}, Adrian},
        title = "{The Manticore Project I: a digital twin of our cosmic neighbourhood from Bayesian field-level analysis}",
      journal = {\mnras},
         year = 2025,
        month = jun,
       volume = {540},
       number = {1},
        pages = {716-745},
          doi = {10.1093/mnras/staf767},
archivePrefix = {arXiv},
       eprint = {2505.10682},
 primaryClass = {astro-ph.CO},
       adsurl = {https://ui.adsabs.harvard.edu/abs/2025MNRAS.540..716M}
}

@ARTICLE{2023MNRAS.526.5682C,
       author = {{Chandran}, Jyothis and {Remazeilles}, Mathieu and {Barreiro}, R.~B.},
        title = "{An improved Compton parameter map of thermal Sunyaev-Zeldovich effect from Planck PR4 data}",
      journal = {\mnras},
         year = 2023,
        month = dec,
       volume = {526},
       number = {4},
        pages = {5682-5698},
          doi = {10.1093/mnras/stad3156},
archivePrefix = {arXiv},
       eprint = {2305.10193},
 primaryClass = {astro-ph.CO},
       adsurl = {https://ui.adsabs.harvard.edu/abs/2023MNRAS.526.5682C}
}

@ARTICLE{2024A&A...685A.106B,
       author = {{Bulbul}, E. and {Liu}, A. and {Kluge}, M. and {Zhang}, X. and {Sanders}, J.~S. and {Bahar}, Y.~E. and {Ghirardini}, V. and {Artis}, E. and {Seppi}, R. and {Garrel}, C. and {Ramos-Ceja}, M.~E. and {Comparat}, J. and {Balzer}, F. and {B{\"o}ckmann}, K. and {Br{\"u}ggen}, M. and {Clerc}, N. and {Dennerl}, K. and {Dolag}, K. and {Freyberg}, M. and {Grandis}, S. and {Gruen}, D. and {Kleinebreil}, F. and {Krippendorf}, S. and {Lamer}, G. and {Merloni}, A. and {Migkas}, K. and {Nandra}, K. and {Pacaud}, F. and {Predehl}, P. and {Reiprich}, T.~H. and {Schrabback}, T. and {Veronica}, A. and {Weller}, J. and {Zelmer}, S.},
        title = "{The SRG/eROSITA All-Sky Survey. The first catalog of galaxy clusters and groups in the Western Galactic Hemisphere}",
      journal = {\aap},
         year = 2024,
        month = may,
       volume = {685},
          eid = {A106},
        pages = {A106},
          doi = {10.1051/0004-6361/202348264},
archivePrefix = {arXiv},
       eprint = {2402.08452},
 primaryClass = {astro-ph.CO},
       adsurl = {https://ui.adsabs.harvard.edu/abs/2024A&A...685A.106B}
}

@ARTICLE{2024A&A...688A.187S,
       author = {{Sadibekova}, T. and {Arnaud}, M. and {Pratt}, G.~W. and {Tarr{\'\i}o}, P. and {Melin}, J. -B.},
        title = "{MCXC-II: Second release of the Meta-Catalogue of X-ray detected Clusters of galaxies}",
      journal = {\aap},
         year = 2024,
        month = aug,
       volume = {688},
          eid = {A187},
        pages = {A187},
          doi = {10.1051/0004-6361/202449427},
archivePrefix = {arXiv},
       eprint = {2402.01538},
 primaryClass = {astro-ph.CO},
       adsurl = {https://ui.adsabs.harvard.edu/abs/2024A&A...688A.187S}
}

@ARTICLE{2025arXiv250206932F,
       author = {{Forouhar Moreno}, Victor J. and {Helly}, John and {McGibbon}, Rob and {Schaye}, Joop and {Schaller}, Matthieu and {Han}, Jiaxin and {Kugel}, Roi},
        title = "{Assessing subhalo finders in cosmological hydrodynamical simulations}",
      journal = {arXiv e-prints},
         year = 2025,
        month = feb,
          eid = {arXiv:2502.06932},
        pages = {arXiv:2502.06932},
          doi = {10.48550/arXiv.2502.06932},
archivePrefix = {arXiv},
       eprint = {2502.06932},
 primaryClass = {astro-ph.CO},
       adsurl = {https://ui.adsabs.harvard.edu/abs/2025arXiv250206932F}
}

@ARTICLE{2025arXiv250722669M,
       author = {{McGibbon}, Robert and {Helly}, John C. and {Schaye}, Joop and {Schaller}, Matthieu and {Vandenbroucke}, Bert},
        title = "{SOAP: A Python Package for Calculating the Properties of Galaxies and Halos Formed in Cosmological Simulations}",
      journal = {arXiv e-prints},
         year = 2025,
        month = jul,
          eid = {arXiv:2507.22669},
        pages = {arXiv:2507.22669},
archivePrefix = {arXiv},
       eprint = {2507.22669},
 primaryClass = {astro-ph.IM},
       adsurl = {https://ui.adsabs.harvard.edu/abs/2025arXiv250722669M}
}

@inproceedings{10.5555/3001460.3001507,
author = {Ester, Martin and Kriegel, Hans-Peter and Sander, J\"{o}rg and Xu, Xiaowei},
title = {A density-based algorithm for discovering clusters in large spatial databases with noise},
year = {1996},
publisher = {AAAI Press},
booktitle = {Proceedings of the Second International Conference on Knowledge Discovery and Data Mining},
pages = {226–231},
numpages = {6},
location = {Portland, Oregon},
series = {KDD'96}
}

@ARTICLE{2018MNRAS.473.1195L,
       author = {{Libeskind}, Noam I. and {van de Weygaert}, Rien and {Cautun}, Marius and {Falck}, Bridget and {Tempel}, Elmo and {Abel}, Tom and {Alpaslan}, Mehmet and {Arag{\'o}n-Calvo}, Miguel A. and {Forero-Romero}, Jaime E. and {Gonzalez}, Roberto and {Gottl{\"o}ber}, Stefan and {Hahn}, Oliver and {Hellwing}, Wojciech A. and {Hoffman}, Yehuda and {Jones}, Bernard J.~T. and {Kitaura}, Francisco and {Knebe}, Alexander and {Manti}, Serena and {Neyrinck}, Mark and {Nuza}, Sebasti{\'a}n E. and {Padilla}, Nelson and {Platen}, Erwin and {Ramachandra}, Nesar and {Robotham}, Aaron and {Saar}, Enn and {Shandarin}, Sergei and {Steinmetz}, Matthias and {Stoica}, Radu S. and {Sousbie}, Thierry and {Yepes}, Gustavo},
        title = "{Tracing the cosmic web}",
      journal = {\mnras},
         year = 2018,
        month = jan,
       volume = {473},
       number = {1},
        pages = {1195-1217},
          doi = {10.1093/mnras/stx1976},
archivePrefix = {arXiv},
       eprint = {1705.03021},
 primaryClass = {astro-ph.CO},
       adsurl = {https://ui.adsabs.harvard.edu/abs/2018MNRAS.473.1195L}
}

@ARTICLE{2014MNRAS.443.1090F,
       author = {{Forero-Romero}, Jaime E. and {Contreras}, Sergio and {Padilla}, Nelson},
        title = "{Cosmic web alignments with the shape, angular momentum and peculiar velocities of dark matter haloes}",
      journal = {\mnras},
         year = 2014,
        month = sep,
       volume = {443},
       number = {2},
        pages = {1090-1102},
          doi = {10.1093/mnras/stu1150},
archivePrefix = {arXiv},
       eprint = {1406.0508},
 primaryClass = {astro-ph.CO},
       adsurl = {https://ui.adsabs.harvard.edu/abs/2014MNRAS.443.1090F}
}

@ARTICLE{2006MNRAS.369.2013R,
       author = {{Rasia}, E. and {Ettori}, S. and {Moscardini}, L. and {Mazzotta}, P. and {Borgani}, S. and {Dolag}, K. and {Tormen}, G. and {Cheng}, L.~M. and {Diaferio}, A.},
        title = "{Systematics in the X-ray cluster mass estimators}",
      journal = {\mnras},
         year = 2006,
        month = jul,
       volume = {369},
       number = {4},
        pages = {2013-2024},
          doi = {10.1111/j.1365-2966.2006.10466.x},
archivePrefix = {arXiv},
       eprint = {astro-ph/0602434},
 primaryClass = {astro-ph},
       adsurl = {https://ui.adsabs.harvard.edu/abs/2006MNRAS.369.2013R}
}

@ARTICLE{2007ApJ...655...98N,
       author = {{Nagai}, Daisuke and {Vikhlinin}, Alexey and {Kravtsov}, Andrey V.},
        title = "{Testing X-Ray Measurements of Galaxy Clusters with Cosmological Simulations}",
      journal = {\apj},
         year = 2007,
        month = jan,
       volume = {655},
       number = {1},
        pages = {98-108},
          doi = {10.1086/509868},
archivePrefix = {arXiv},
       eprint = {astro-ph/0609247},
 primaryClass = {astro-ph},
       adsurl = {https://ui.adsabs.harvard.edu/abs/2007ApJ...655...98N}
}

@ARTICLE{2024MNRAS.534..251K,
       author = {{Kay}, Scott T. and {Braspenning}, Joey and {Chluba}, Jens and {Helly}, John C. and {Kugel}, Roi and {Schaller}, Matthieu and {Schaye}, Joop},
        title = "{Relativistic SZ temperatures and hydrostatic mass bias for massive clusters in the FLAMINGO simulations}",
      journal = {\mnras},
         year = 2024,
        month = oct,
       volume = {534},
       number = {1},
        pages = {251-270},
          doi = {10.1093/mnras/stae1991},
archivePrefix = {arXiv},
       eprint = {2404.08539},
 primaryClass = {astro-ph.CO},
       adsurl = {https://ui.adsabs.harvard.edu/abs/2024MNRAS.534..251K}
}

@ARTICLE{2016A&A...594A..24P,
       author = {{Planck Collaboration} and {Ade}, P.~A.~R. and {Aghanim}, N. and {Arnaud}, M. and {Ashdown}, M. and {Aumont}, J. and {Baccigalupi}, C. and {Banday}, A.~J. and {Barreiro}, R.~B. and {Bartlett}, J.~G. and {Bartolo}, N. and {Battaner}, E. and {Battye}, R. and {Benabed}, K. and {Beno{\^\i}t}, A. and {Benoit-L{\'e}vy}, A. and {Bernard}, J. -P. and {Bersanelli}, M. and {Bielewicz}, P. and {Bock}, J.~J. and {Bonaldi}, A. and {Bonavera}, L. and {Bond}, J.~R. and {Borrill}, J. and {Bouchet}, F.~R. and {Bucher}, M. and {Burigana}, C. and {Butler}, R.~C. and {Calabrese}, E. and {Cardoso}, J. -F. and {Catalano}, A. and {Challinor}, A. and {Chamballu}, A. and {Chary}, R. -R. and {Chiang}, H.~C. and {Christensen}, P.~R. and {Church}, S. and {Clements}, D.~L. and {Colombi}, S. and {Colombo}, L.~P.~L. and {Combet}, C. and {Comis}, B. and {Couchot}, F. and {Coulais}, A. and {Crill}, B.~P. and {Curto}, A. and {Cuttaia}, F. and {Danese}, L. and {Davies}, R.~D. and {Davis}, R.~J. and {de Bernardis}, P. and {de Rosa}, A. and {de Zotti}, G. and {Delabrouille}, J. and {D{\'e}sert}, F. -X. and {Diego}, J.~M. and {Dolag}, K. and {Dole}, H. and {Donzelli}, S. and {Dor{\'e}}, O. and {Douspis}, M. and {Ducout}, A. and {Dupac}, X. and {Efstathiou}, G. and {Elsner}, F. and {En{\ss}lin}, T.~A. and {Eriksen}, H.~K. and {Falgarone}, E. and {Fergusson}, J. and {Finelli}, F. and {Forni}, O. and {Frailis}, M. and {Fraisse}, A.~A. and {Franceschi}, E. and {Frejsel}, A. and {Galeotta}, S. and {Galli}, S. and {Ganga}, K. and {Giard}, M. and {Giraud-H{\'e}raud}, Y. and {Gjerl{\o}w}, E. and {Gonz{\'a}lez-Nuevo}, J. and {G{\'o}rski}, K.~M. and {Gratton}, S. and {Gregorio}, A. and {Gruppuso}, A. and {Gudmundsson}, J.~E. and {Hansen}, F.~K. and {Hanson}, D. and {Harrison}, D.~L. and {Henrot-Versill{\'e}}, S. and {Hern{\'a}ndez-Monteagudo}, C. and {Herranz}, D. and {Hildebrandt}, S.~R. and {Hivon}, E. and {Hobson}, M. and {Holmes}, W.~A. and {Hornstrup}, A. and {Hovest}, W. and {Huffenberger}, K.~M. and {Hurier}, G. and {Jaffe}, A.~H. and {Jaffe}, T.~R. and {Jones}, W.~C. and {Juvela}, M. and {Keih{\"a}nen}, E. and {Keskitalo}, R. and {Kisner}, T.~S. and {Kneissl}, R. and {Knoche}, J. and {Kunz}, M. and {Kurki-Suonio}, H. and {Lagache}, G. and {L{\"a}hteenm{\"a}ki}, A. and {Lamarre}, J. -M. and {Lasenby}, A. and {Lattanzi}, M. and {Lawrence}, C.~R. and {Leonardi}, R. and {Lesgourgues}, J. and {Levrier}, F. and {Liguori}, M. and {Lilje}, P.~B. and {Linden-V{\o}rnle}, M. and {L{\'o}pez-Caniego}, M. and {Lubin}, P.~M. and {Mac{\'\i}as-P{\'e}rez}, J.~F. and {Maggio}, G. and {Maino}, D. and {Mandolesi}, N. and {Mangilli}, A. and {Maris}, M. and {Martin}, P.~G. and {Mart{\'\i}nez-Gonz{\'a}lez}, E. and {Masi}, S. and {Matarrese}, S. and {McGehee}, P. and {Meinhold}, P.~R. and {Melchiorri}, A. and {Melin}, J. -B. and {Mendes}, L. and {Mennella}, A. and {Migliaccio}, M. and {Mitra}, S. and {Miville-Desch{\^e}nes}, M. -A. and {Moneti}, A. and {Montier}, L. and {Morgante}, G. and {Mortlock}, D. and {Moss}, A. and {Munshi}, D. and {Murphy}, J.~A. and {Naselsky}, P. and {Nati}, F. and {Natoli}, P. and {Netterfield}, C.~B. and {N{\o}rgaard-Nielsen}, H.~U. and {Noviello}, F. and {Novikov}, D. and {Novikov}, I. and {Oxborrow}, C.~A. and {Paci}, F. and {Pagano}, L. and {Pajot}, F. and {Paoletti}, D. and {Partridge}, B. and {Pasian}, F. and {Patanchon}, G. and {Pearson}, T.~J. and {Perdereau}, O. and {Perotto}, L. and {Perrotta}, F. and {Pettorino}, V. and {Piacentini}, F. and {Piat}, M. and {Pierpaoli}, E. and {Pietrobon}, D. and {Plaszczynski}, S. and {Pointecouteau}, E. and {Polenta}, G. and {Popa}, L. and {Pratt}, G.~W. and {Pr{\'e}zeau}, G. and {Prunet}, S. and {Puget}, J. -L. and {Rachen}, J.~P. and {Rebolo}, R. and {Reinecke}, M. and {Remazeilles}, M. and {Renault}, C. and {Renzi}, A. and {Ristorcelli}, I. and {Rocha}, G. and {Roman}, M. and {Rosset}, C. and {Rossetti}, M. and {Roudier}, G. and {Rubi{\~n}o-Mart{\'\i}n}, J.~A. and {Rusholme}, B. and {Sandri}, M.},
        title = "{Planck 2015 results. XXIV. Cosmology from Sunyaev-Zeldovich cluster counts}",
      journal = {\aap},
         year = 2016,
        month = sep,
       volume = {594},
          eid = {A24},
        pages = {A24},
          doi = {10.1051/0004-6361/201525833},
archivePrefix = {arXiv},
       eprint = {1502.01597},
 primaryClass = {astro-ph.CO},
       adsurl = {https://ui.adsabs.harvard.edu/abs/2016A&A...594A..24P}
}

@ARTICLE{2024A&A...689A.298G,
       author = {{Ghirardini}, V. and {Bulbul}, E. and {Artis}, E. and {Clerc}, N. and {Garrel}, C. and {Grandis}, S. and {Kluge}, M. and {Liu}, A. and {Bahar}, Y.~E. and {Balzer}, F. and {Chiu}, I. and {Comparat}, J. and {Gruen}, D. and {Kleinebreil}, F. and {Krippendorf}, S. and {Merloni}, A. and {Nandra}, K. and {Okabe}, N. and {Pacaud}, F. and {Predehl}, P. and {Ramos-Ceja}, M.~E. and {Reiprich}, T.~H. and {Sanders}, J.~S. and {Schrabback}, T. and {Seppi}, R. and {Zelmer}, S. and {Zhang}, X. and {Bornemann}, W. and {Brunner}, H. and {Burwitz}, V. and {Coutinho}, D. and {Dennerl}, K. and {Freyberg}, M. and {Friedrich}, S. and {Gaida}, R. and {Gueguen}, A. and {Haberl}, F. and {Kink}, W. and {Lamer}, G. and {Li}, X. and {Liu}, T. and {Maitra}, C. and {Meidinger}, N. and {Mueller}, S. and {Miyatake}, H. and {Miyazaki}, S. and {Robrade}, J. and {Schwope}, A. and {Stewart}, I.},
        title = "{The SRG/eROSITA all-sky survey: Cosmology constraints from cluster abundances in the western Galactic hemisphere}",
      journal = {\aap},
         year = 2024,
        month = sep,
       volume = {689},
          eid = {A298},
        pages = {A298},
          doi = {10.1051/0004-6361/202348852},
archivePrefix = {arXiv},
       eprint = {2402.08458},
 primaryClass = {astro-ph.CO},
       adsurl = {https://ui.adsabs.harvard.edu/abs/2024A&A...689A.298G}
}

@ARTICLE{2021MNRAS.506.2533B,
       author = {{Barnes}, David J. and {Vogelsberger}, Mark and {Pearce}, Francesca A. and {Pop}, Ana-Roxana and {Kannan}, Rahul and {Cao}, Kaili and {Kay}, Scott T. and {Hernquist}, Lars},
        title = "{Characterizing hydrostatic mass bias with MOCK-X}",
      journal = {\mnras},
         year = 2021,
        month = sep,
       volume = {506},
       number = {2},
        pages = {2533-2550},
          doi = {10.1093/mnras/stab1276},
archivePrefix = {arXiv},
       eprint = {2001.11508},
 primaryClass = {astro-ph.CO},
       adsurl = {https://ui.adsabs.harvard.edu/abs/2021MNRAS.506.2533B}
}

@ARTICLE{2011ARA&A..49..409A,
       author = {{Allen}, Steven W. and {Evrard}, August E. and {Mantz}, Adam B.},
        title = "{Cosmological Parameters from Observations of Galaxy Clusters}",
      journal = {\araa},
         year = 2011,
        month = sep,
       volume = {49},
       number = {1},
        pages = {409-470},
          doi = {10.1146/annurev-astro-081710-102514},
archivePrefix = {arXiv},
       eprint = {1103.4829},
 primaryClass = {astro-ph.CO},
       adsurl = {https://ui.adsabs.harvard.edu/abs/2011ARA&A..49..409A}
}

@ARTICLE{2012ARA&A..50..353K,
       author = {{Kravtsov}, Andrey V. and {Borgani}, Stefano},
        title = "{Formation of Galaxy Clusters}",
      journal = {\araa},
         year = 2012,
        month = sep,
       volume = {50},
        pages = {353-409},
          doi = {10.1146/annurev-astro-081811-125502},
archivePrefix = {arXiv},
       eprint = {1205.5556},
 primaryClass = {astro-ph.CO},
       adsurl = {https://ui.adsabs.harvard.edu/abs/2012ARA&A..50..353K}
}

@ARTICLE{2004ApJ...606..702T,
       author = {{Tegmark}, Max and {Blanton}, Michael R. and {Strauss}, Michael A. and {Hoyle}, Fiona and {Schlegel}, David and {Scoccimarro}, Roman and {Vogeley}, Michael S. and {Weinberg}, David H. and {Zehavi}, Idit and {Berlind}, Andreas and {Budavari}, Tam{\'a}s and {Connolly}, Andrew and {Eisenstein}, Daniel J. and {Finkbeiner}, Douglas and {Frieman}, Joshua A. and {Gunn}, James E. and {Hamilton}, Andrew J.~S. and {Hui}, Lam and {Jain}, Bhuvnesh and {Johnston}, David and {Kent}, Stephen and {Lin}, Huan and {Nakajima}, Reiko and {Nichol}, Robert C. and {Ostriker}, Jeremiah P. and {Pope}, Adrian and {Scranton}, Ryan and {Seljak}, Uro{\v{s}} and {Sheth}, Ravi K. and {Stebbins}, Albert and {Szalay}, Alexander S. and {Szapudi}, Istv{\'a}n and {Verde}, Licia and {Xu}, Yongzhong and {Annis}, James and {Bahcall}, Neta A. and {Brinkmann}, J. and {Burles}, Scott and {Castander}, Francisco J. and {Csabai}, Istvan and {Loveday}, Jon and {Doi}, Mamoru and {Fukugita}, Masataka and {Gott}, III, J. Richard and {Hennessy}, Greg and {Hogg}, David W. and {Ivezi{\'c}}, {\v{Z}}eljko and {Knapp}, Gillian R. and {Lamb}, Don Q. and {Lee}, Brian C. and {Lupton}, Robert H. and {McKay}, Timothy A. and {Kunszt}, Peter and {Munn}, Jeffrey A. and {O'Connell}, Liam and {Peoples}, John and {Pier}, Jeffrey R. and {Richmond}, Michael and {Rockosi}, Constance and {Schneider}, Donald P. and {Stoughton}, Christopher and {Tucker}, Douglas L. and {Vanden Berk}, Daniel E. and {Yanny}, Brian and {York}, Donald G. and {SDSS Collaboration}},
        title = "{The Three-Dimensional Power Spectrum of Galaxies from the Sloan Digital Sky Survey}",
      journal = {\apj},
         year = 2004,
        month = may,
       volume = {606},
       number = {2},
        pages = {702-740},
          doi = {10.1086/382125},
archivePrefix = {arXiv},
       eprint = {astro-ph/0310725},
 primaryClass = {astro-ph},
       adsurl = {https://ui.adsabs.harvard.edu/abs/2004ApJ...606..702T}
}

@ARTICLE{2005MNRAS.362..505C,
       author = {{Cole}, Shaun and {Percival}, Will J. and {Peacock}, John A. and {Norberg}, Peder and {Baugh}, Carlton M. and {Frenk}, Carlos S. and {Baldry}, Ivan and {Bland-Hawthorn}, Joss and {Bridges}, Terry and {Cannon}, Russell and {Colless}, Matthew and {Collins}, Chris and {Couch}, Warrick and {Cross}, Nicholas J.~G. and {Dalton}, Gavin and {Eke}, Vincent R. and {De Propris}, Roberto and {Driver}, Simon P. and {Efstathiou}, George and {Ellis}, Richard S. and {Glazebrook}, Karl and {Jackson}, Carole and {Jenkins}, Adrian and {Lahav}, Ofer and {Lewis}, Ian and {Lumsden}, Stuart and {Maddox}, Steve and {Madgwick}, Darren and {Peterson}, Bruce A. and {Sutherland}, Will and {Taylor}, Keith},
        title = "{The 2dF Galaxy Redshift Survey: power-spectrum analysis of the final data set and cosmological implications}",
      journal = {\mnras},
         year = 2005,
        month = sep,
       volume = {362},
       number = {2},
        pages = {505-534},
          doi = {10.1111/j.1365-2966.2005.09318.x},
archivePrefix = {arXiv},
       eprint = {astro-ph/0501174},
 primaryClass = {astro-ph},
       adsurl = {https://ui.adsabs.harvard.edu/abs/2005MNRAS.362..505C}
}

@ARTICLE{2022A&A...661A.146B,
       author = {{Bonnaire}, Tony and {Aghanim}, Nabila and {Kuruvilla}, Joseph and {Decelle}, Aur{\'e}lien},
        title = "{Cosmology with cosmic web environments. I. Real-space power spectra}",
      journal = {\aap},
         year = 2022,
        month = may,
       volume = {661},
          eid = {A146},
        pages = {A146},
          doi = {10.1051/0004-6361/202142852},
archivePrefix = {arXiv},
       eprint = {2112.03926},
 primaryClass = {astro-ph.CO},
       adsurl = {https://ui.adsabs.harvard.edu/abs/2022A&A...661A.146B}
}

@ARTICLE{2013ApJ...762..115O,
       author = {{Obreschkow}, D. and {Power}, C. and {Bruderer}, M. and {Bonvin}, C.},
        title = "{A Robust Measure of Cosmic Structure beyond the Power Spectrum: Cosmic Filaments and the Temperature of Dark Matter}",
      journal = {\apj},
         year = 2013,
        month = jan,
       volume = {762},
       number = {2},
          eid = {115},
        pages = {115},
          doi = {10.1088/0004-637X/762/2/115},
archivePrefix = {arXiv},
       eprint = {1211.5213},
 primaryClass = {astro-ph.CO},
       adsurl = {https://ui.adsabs.harvard.edu/abs/2013ApJ...762..115O}
}

@ARTICLE{2025arXiv250706866M,
       author = {{Malandrino}, Rosa and {Lavaux}, Guilhem and {Wandelt}, Benjamin D. and {McAlpine}, Stuart and {Jasche}, Jens},
        title = "{A Bayesian catalog of 100 high-significance voids in the Local Universe}",
      journal = {arXiv e-prints},
         year = 2025,
        month = jul,
          eid = {arXiv:2507.06866},
        pages = {arXiv:2507.06866},
          doi = {10.48550/arXiv.2507.06866},
archivePrefix = {arXiv},
       eprint = {2507.06866},
 primaryClass = {astro-ph.CO},
       adsurl = {https://ui.adsabs.harvard.edu/abs/2025arXiv250706866M}
}

@INPROCEEDINGS{Bos2016,
       author = {{Bos}, E.~G.~P. and {van de Weygaert}, Rien and {Kitaura}, Francisco-Shu and {Cautun}, Marius},
        title = "{Bayesian Cosmic Web Reconstruction: BARCODE for Clusters}",
    booktitle = {The Zeldovich Universe: Genesis and Growth of the Cosmic Web},
         year = 2016,
       editor = {{van de Weygaert}, Rien and {Shandarin}, Sergei and {Saar}, Enn and {Einasto}, Jaan},
       series = {IAU Symposium},
       volume = {308},
        month = oct,
        pages = {271-288},
          doi = {10.1017/S1743921316009996},
archivePrefix = {arXiv},
       eprint = {1611.01220},
 primaryClass = {astro-ph.CO},
       adsurl = {https://ui.adsabs.harvard.edu/abs/2016IAUS..308..271B}
}

@ARTICLE{Kitaura2021,
       author = {{Kitaura}, Francisco-Shu and {Ata}, Metin and {Rodriguez-Torres}, Sergio A. and {Hernandez-Sanchez}, Monica and {Balaguera-Antolinez}, A. and {Yepes}, Gustavo},
        title = "{cosmic birth: efficient Bayesian inference of the evolving cosmic web from galaxy surveys}",
      journal = {\mnras},
         year = 2021,
        month = apr,
       volume = {502},
       number = {3},
        pages = {3456-3475},
          doi = {10.1093/mnras/staa3774},
archivePrefix = {arXiv},
       eprint = {1911.00284},
 primaryClass = {astro-ph.CO},
       adsurl = {https://ui.adsabs.harvard.edu/abs/2021MNRAS.502.3456K}
}

@ARTICLE{Sarpa2022,
       author = {{Sarpa}, E. and {Longobardi}, A. and {Kraljic}, K. and {Veropalumbo}, A. and {Schimd}, C.},
        title = "{Tracing the environmental history of observed galaxies via extended fast action minimization method}",
      journal = {\mnras},
         year = 2022,
        month = oct,
       volume = {516},
       number = {1},
        pages = {231-244},
          doi = {10.1093/mnras/stac2125},
archivePrefix = {arXiv},
       eprint = {2204.09709},
 primaryClass = {astro-ph.CO},
       adsurl = {https://ui.adsabs.harvard.edu/abs/2022MNRAS.516..231S}
}

@ARTICLE{Veena2023,
       author = {{Ganeshaiah Veena}, Punyakoti and {Lilow}, Robert and {Nusser}, Adi},
        title = "{Large-scale density and velocity field reconstructions with neural networks}",
      journal = {\mnras},
         year = 2023,
        month = jul,
       volume = {522},
       number = {4},
        pages = {5291-5307},
          doi = {10.1093/mnras/stad1222},
archivePrefix = {arXiv},
       eprint = {2302.14101},
 primaryClass = {astro-ph.CO},
       adsurl = {https://ui.adsabs.harvard.edu/abs/2023MNRAS.522.5291G}
}

@ARTICLE{Boselli2014,
       author = {{Boselli}, A. and {Voyer}, E. and {Boissier}, S. and {Cucciati}, O. and {Consolandi}, G. and {Cortese}, L. and {Fumagalli}, M. and {Gavazzi}, G. and {Heinis}, S. and {Roehlly}, Y.},
        title = "{The GALEX Ultraviolet Virgo Cluster Survey (GUViCS). IV. The role of the cluster environment on galaxy evolution}",
      journal = {\aap},
         year = 2014,
        month = oct,
       volume = {570},
          eid = {A69},
        pages = {A69},
          doi = {10.1051/0004-6361/201424419},
archivePrefix = {arXiv},
       eprint = {1407.6990},
 primaryClass = {astro-ph.GA},
       adsurl = {https://ui.adsabs.harvard.edu/abs/2014A&A...570A..69B}
}

@ARTICLE{Kubo2007,
       author = {{Kubo}, Jeffrey M. and {Stebbins}, Albert and {Annis}, James and {Dell'Antonio}, Ian P. and {Lin}, Huan and {Khiabanian}, Hossein and {Frieman}, Joshua A.},
        title = "{The Mass of the Coma Cluster from Weak Lensing in the Sloan Digital Sky Survey}",
      journal = {\apj},
         year = 2007,
        month = dec,
       volume = {671},
       number = {2},
        pages = {1466-1470},
          doi = {10.1086/523101},
archivePrefix = {arXiv},
       eprint = {0709.0506},
 primaryClass = {astro-ph},
       adsurl = {https://ui.adsabs.harvard.edu/abs/2007ApJ...671.1466K}
}

@ARTICLE{Springel2005,
       author = {{Springel}, Volker and {White}, Simon D.~M. and {Jenkins}, Adrian and {Frenk}, Carlos S. and {Yoshida}, Naoki and {Gao}, Liang and {Navarro}, Julio and {Thacker}, Robert and {Croton}, Darren and {Helly}, John and {Peacock}, John A. and {Cole}, Shaun and {Thomas}, Peter and {Couchman}, Hugh and {Evrard}, August and {Colberg}, J{\"o}rg and {Pearce}, Frazer},
        title = "{Simulations of the formation, evolution and clustering of galaxies and quasars}",
      journal = {\nat},
         year = 2005,
        month = jun,
       volume = {435},
       number = {7042},
        pages = {629-636},
          doi = {10.1038/nature03597},
archivePrefix = {arXiv},
       eprint = {astro-ph/0504097},
 primaryClass = {astro-ph},
       adsurl = {https://ui.adsabs.harvard.edu/abs/2005Natur.435..629S}
}

@ARTICLE{Klypin2016,
       author = {{Klypin}, Anatoly and {Yepes}, Gustavo and {Gottl{\"o}ber}, Stefan and {Prada}, Francisco and {He{\ss}}, Steffen},
        title = "{MultiDark simulations: the story of dark matter halo concentrations and density profiles}",
      journal = {\mnras},
         year = 2016,
        month = apr,
       volume = {457},
       number = {4},
        pages = {4340-4359},
          doi = {10.1093/mnras/stw248},
archivePrefix = {arXiv},
       eprint = {1411.4001},
 primaryClass = {astro-ph.CO},
       adsurl = {https://ui.adsabs.harvard.edu/abs/2016MNRAS.457.4340K}
}

@ARTICLE{Nelson2019,
       author = {{Nelson}, Dylan and {Springel}, Volker and {Pillepich}, Annalisa and {Rodriguez-Gomez}, Vicente and {Torrey}, Paul and {Genel}, Shy and {Vogelsberger}, Mark and {Pakmor}, R{\"u}diger and {Marinacci}, Federico and {Weinberger}, Rainer and {Kelley}, Luke and {Lovell}, Mark and {Diemer}, Benedikt and {Hernquist}, Lars},
        title = "{The IllustrisTNG simulations: public data release}",
      journal = {Computational Astrophysics and Cosmology},
         year = 2019,
        month = mar,
       volume = {6},
       number = {1},
          eid = {2},
        pages = {2},
          doi = {10.1186/s40668-019-0028-x},
archivePrefix = {arXiv},
       eprint = {1812.05609},
 primaryClass = {astro-ph.GA},
       adsurl = {https://ui.adsabs.harvard.edu/abs/2019ComAC...6....2N}
}

\appendix
\section{DBSCAN Parameter Selection and Catalogue Robustness}
\label{app:dbscan}

This appendix describes the parameter selection and robustness analysis for the DBSCAN clustering methodology used to identify posterior halo associations. We demonstrate that the recovered associations are insensitive to reasonable parameter choices and present the selection criteria used to construct the final catalogue. To avoid confusion with galaxy clusters in the cosmological sense, we use the term \textit{DBSCAN cluster} to refer to the raw groupings identified by the algorithm, and \textit{posterior association} to refer to the physical structures in the final catalogue after applying the one-per-realization constraint and quality cuts.

As a reminder, the clustering procedure operates on the combined halo catalogue formed by stacking central haloes from all posterior realizations. DBSCAN groups haloes based on spatial proximity, controlled by two parameters: the neighbourhood radius, $\epsilon$, and the minimum number of neighbours required to seed a DBSCAN cluster, $N_{\rm min}$. A halo is classified as a core point if at least $N_{\rm min}$ other haloes lie within distance $\epsilon$; DBSCAN clusters grow by linking core points and their neighbours into connected components. Haloes that belong to no DBSCAN cluster are labelled as noise.

Because each realization represents an independent draw from the Bayesian posterior, a single physical structure can contribute at most one halo per realization. We therefore enforce a one-per-realization constraint: when DBSCAN initially assigns multiple haloes from the same realization to a DBSCAN cluster, we retain only the halo closest to the centroid. This constraint is physically motivated, a massive halo cannot appear twice in a single realization, and ensures that association membership counts directly reflect the posterior probability that a structure exists. The fraction of realizations in which this constraint must be applied defines the \textit{ambiguity rate} of a posterior association; lower values indicate cleaner initial assignments with fewer conflicts to resolve.

Multi-candidate situations arise when DBSCAN merges haloes from physically adjacent structures that happen to fall within the neighbourhood radius $\epsilon$. This typically occurs in crowded environments where two or more massive haloes at similar distances could plausibly satisfy the same observational constraints, or when $\epsilon$ is set large enough that genuinely distinct structures are grouped together. By selecting the halo closest to the centroid, we adopt a conservative approach that favours the most centrally located candidate within each realization. This choice may introduce a modest bias toward more centrally concentrated configurations, potentially underestimating the true spatial extent of associations in crowded regions. However, the alternative of random selection or mass-weighted selection would introduce different biases, and the centroid-based approach has the advantage of being deterministic and reproducible. Importantly, associations with high ambiguity rates are flagged and can be excluded via quality cuts, ensuring that the final catalogue prioritizes clean assignments where this selection rarely occurs.

\begin{figure*}
    \includegraphics[width=\textwidth]{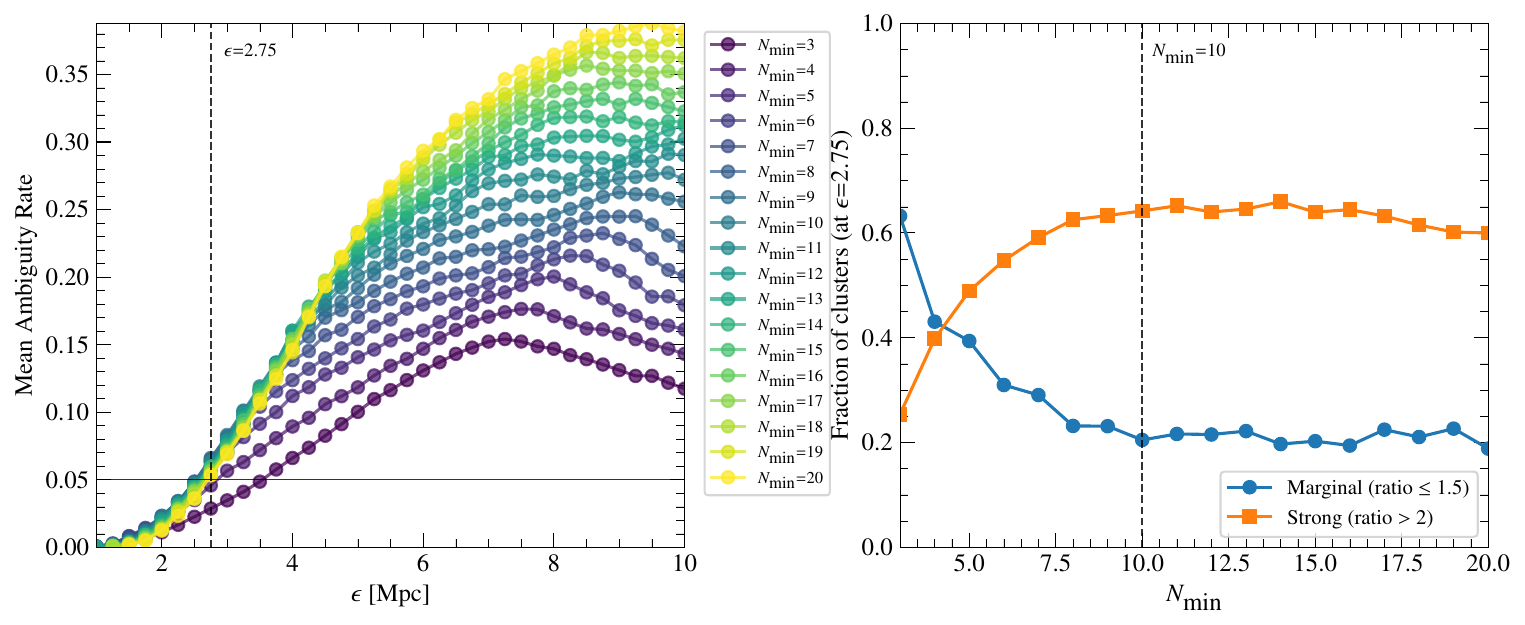}
    \caption{Fiducial parameter selection for DBSCAN clustering. \textit{Left}: Mean ambiguity rate as a function of the neighbourhood radius $\epsilon$ for different values of $N_{\rm min}$ (coloured lines). The transition from no ambiguity to some ambiguity occurs at approximately the same $\epsilon$ regardless of $N_{\rm min}$, indicating that the onset of over-merging is primarily controlled by $\epsilon$. The vertical dashed line marks $\epsilon = 2.75$~Mpc, where the mean ambiguity rate of DBSCAN clusters crosses the 5\% threshold. \textit{Right}: Fraction of DBSCAN clusters classified as marginal ($N_{\rm members}/N_{\rm min} < 1.5$, blue) or strong ($N_{\rm members}/N_{\rm min} > 2$, orange) detections as a function of $N_{\rm min}$ at fixed $\epsilon = 2.75$~Mpc. The strong fraction plateaus at $N_{\rm min} \approx 10$, indicating that further increases filter noise without improving the purity of genuine detections.}
    \label{fig:dbscan_sensitivity}
\end{figure*}

To assess the sensitivity of clustering results to parameter choices and derive principled fiducial values, we perform a systematic sweep over a two-dimensional grid spanning $\epsilon = 1$--$10$~Mpc and $N_{\rm min} = 3$--$20$, yielding approximately 300 parameter combinations. Each combination produces an independent set of posterior association candidates from the same input halo catalogue. We exclude parameter runs that yield fewer than 100 DBSCAN clusters, as these represent phase-transition regimes where the algorithm either fragments the field entirely or merges most haloes into a single structure. Only 9 of the 300 combinations (3\%) fall into this degenerate regime, all at extreme parameter values.

Rather than selecting fiducial parameters by maximizing a single summary statistic, we adopt a two-stage approach that addresses the distinct failure modes of each parameter. Increasing $\epsilon$ beyond the true correlation scale causes over-merging, where physically distinct structures are grouped together; this manifests as elevated ambiguity rates. Conversely, setting $N_{\rm min}$ too low admits noise fluctuations as spurious detections, while setting it too high discards genuine associations with modest membership. We therefore use the ambiguity rate to constrain $\epsilon$ and a membership-ratio diagnostic to constrain $N_{\rm min}$.

The left panel of \cref{fig:dbscan_sensitivity} shows the mean ambiguity rate as a function of $\epsilon$ for each value of $N_{\rm min}$. At small $\epsilon$, the ambiguity rate is near zero because neighbourhood volumes are too small to encompass multiple haloes from the same realization. As $\epsilon$ increases, the rate rises as DBSCAN clusters begin to merge physically adjacent structures. The transition from no ambiguity to some ambiguity occurs at approximately the same $\epsilon$ regardless of $N_{\rm min}$, indicating that the onset of over-merging is primarily controlled by the neighbourhood radius rather than the density threshold. We select $\epsilon = 2.75$~Mpc as the fiducial value, corresponding to the threshold where the mean ambiguity rate of DBSCAN clusters crosses 5\%. Below this scale, the majority of initial DBSCAN assignments are clean and the one-per-realization constraint rarely needs to arbitrate between multiple candidates.

With $\epsilon$ fixed, we turn to $N_{\rm min}$. A DBSCAN cluster that barely exceeds the membership threshold is more likely to represent a noise fluctuation than a genuine structure. We quantify this using the ratio $N_{\rm members}/N_{\rm min}$: clusters with ratios below 1.5 are classified as \textit{marginal} detections, while those exceeding 2 are \textit{strong} detections with membership well above the threshold. The right panel of \cref{fig:dbscan_sensitivity} shows these fractions as a function of $N_{\rm min}$ at the fiducial $\epsilon = 2.75$~Mpc. At low $N_{\rm min}$, the marginal fraction dominates, most detections barely exceed the threshold and are likely spurious. As $N_{\rm min}$ increases, the marginal fraction drops sharply while the strong fraction rises, indicating that noise is being filtered out. The strong fraction plateaus at $N_{\rm min} \approx 10$, beyond which further increases reduce the total cluster count without improving the purity of genuine detections. We therefore adopt $N_{\rm min} = 10$ as the fiducial value.

This two-stage selection yields fiducial parameters $\epsilon = 2.75$~Mpc and $N_{\rm min} = 10$ that are derived from physically motivated diagnostics rather than arbitrary optimization targets. The ambiguity rate constrains $\epsilon$ to avoid over-merging, while the membership-ratio diagnostic constrains $N_{\rm min}$ to filter noise without discarding real associations. At the fiducial parameters, DBSCAN identifies 1,210 DBSCAN clusters prior to any quality cuts. The mean ambiguity rate at these parameters is below 5\%, confirming that the one-per-realization constraint is rarely invoked and the initial assignments are clean.

We note that the properties of the recovered associations---including their spatial coherence, mass distributions, and membership counts---ultimately depend on the assumptions underlying the \manticorelocal\ reconstruction, particularly the galaxy bias model, the dynamical forward model, and the prior on initial conditions. These modelling choices propagate into the posterior halo catalogues that serve as input to the clustering procedure. While the DBSCAN algorithm itself is agnostic to these assumptions, the robustness and physical interpretation of the resulting associations inherit the uncertainties and potential systematics of the parent inference. Users should therefore interpret association properties as posterior summaries conditioned on the \manticorelocal\ framework rather than model-independent measurements.

To verify that the recovered associations represent genuine structures rather than noise artefacts, we apply the DBSCAN pipeline to control simulations with randomized observer positions. Rather than sweeping the full parameter grid, we test control simulations at the fiducial parameters and at extreme values of $\epsilon$ and $N_{\rm min}$. Only at extreme parameter settings, large $\epsilon$ combined with small $N_{\rm min}$, does DBSCAN identify any groupings in the control fields. At our fiducial parameters ($\epsilon = 2.75$~Mpc, $N_{\rm min} = 10$), the algorithm finds zero DBSCAN clusters in the randomized control simulations. This result confirms that the posterior associations identified in the \manticorelocal\ posterior are not chance alignments that would appear in any stacked halo catalogue, but rather reflect genuine spatial correlations imposed by the observational constraints.

\begin{table}
    \centering
    \caption{Posterior association catalogue selection cuts. All selections include a mass cut $\langle M_{200} \rangle \geq 10^{14}\,{\rm M_\odot}$, restricting the catalogue to cluster-scale systems. Each row applies cumulative cuts to the raw DBSCAN output at fiducial parameters ($\epsilon = 2.75$~Mpc, $N_{\rm min} = 10$). The \textbf{Fiducial} sample represents our recommended selection for general analyses.}
    \label{tab:selection_cuts}
    \begin{tabular}{lcccccc}
        \toprule
        Selection
        & $\langle M_{200} \rangle$
        & Ambiguity
        & $N_{\rm members}$
        & $\sigma_R$
        & $\sigma_{\log M}$
        & $N_{\rm assoc}$ \\
        & $[{\rm M_\odot}]$
        & rate
        & 
        & [Mpc]
        & [dex]
        & \\
        \midrule
        Raw
        & $\geq 10^{14}$
        & \multicolumn{1}{c}{--}
        & \multicolumn{1}{c}{--}
        & \multicolumn{1}{c}{--}
        & \multicolumn{1}{c}{--}
        & 1210 \\
        \textbf{Fiducial}
        & $\geq 10^{14}$
        & $< 0.05$
        & $\geq 20$
        & \multicolumn{1}{c}{--}
        & \multicolumn{1}{c}{--}
        & \textbf{401} \\
        Strict
        & $\geq 10^{14}$
        & $< 0.05$
        & $\geq 20$
        & $< 2.5$
        & $< 0.25$
        & 162 \\
        \bottomrule
    \end{tabular}
\end{table}

From the fiducial DBSCAN output, we construct the final posterior association catalogue by applying quality cuts summarized in \cref{tab:selection_cuts}. All selections include a mass cut $\langle M_{200} \rangle \geq 10^{14}\,{\rm M_\odot}$, restricting the catalogue to cluster-scale systems; this study focuses exclusively on associations above this threshold. The \textit{Raw} sample contains all 1,210 associations with $\langle M_{200} \rangle \geq 10^{14}\,{\rm M_\odot}$ identified at the fiducial parameters. The \textit{Fiducial} sample retains 401 associations (33\%) by requiring ambiguity rate below 5\% to select associations with clean initial assignments, and at least 20 member haloes to ensure adequate sampling of the posterior. The membership threshold of 20 is motivated by the need to reliably estimate posterior summary statistics such as the mean and median position and mass; associations with fewer members provide limited constraints on the underlying structure. The \textit{Strict} sample further restricts to 162 associations (13\%) by additionally requiring spatial scatter $\sigma_R < 2.5$~Mpc and mass scatter $\sigma_{\log M} < 0.25$~dex, selecting the most tightly constrained structures for applications requiring robust predictions.

\begin{figure}
    \includegraphics[width=\columnwidth]{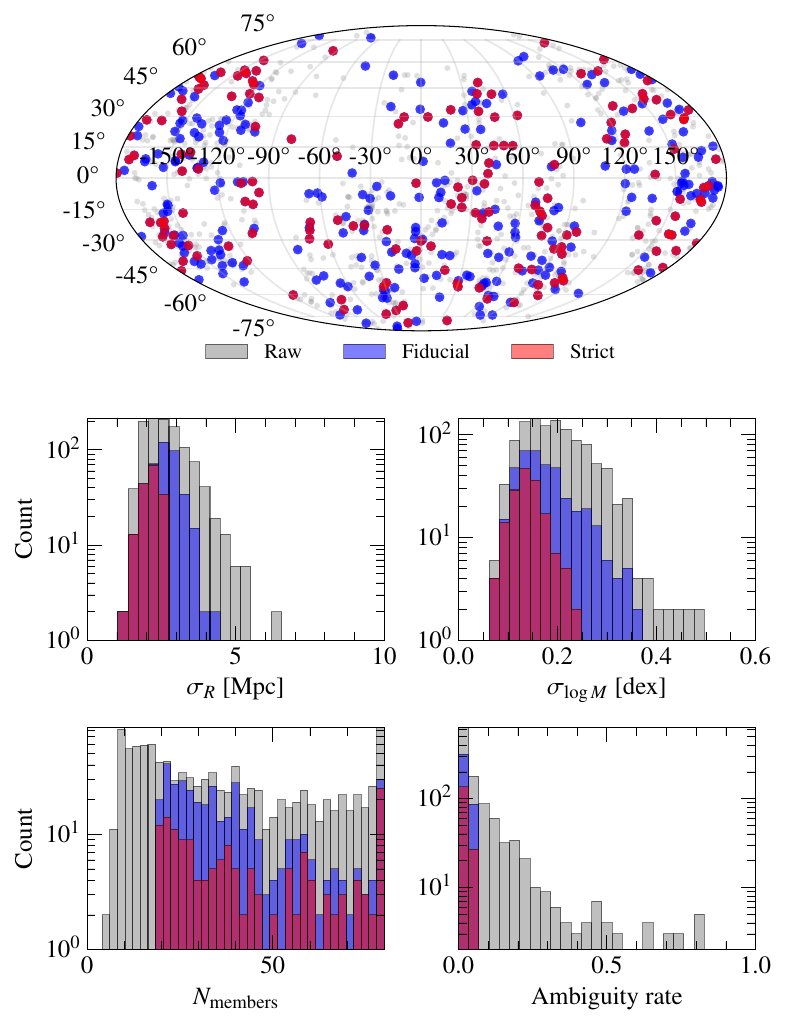}
    \caption{Distribution of posterior association properties under different selection cuts. \textit{Top}: On-sky distribution in Mollweide projection (grey: raw, blue: fiducial, red: strict). \textit{Middle row}: Spatial scatter $\sigma_R$ (left) and mass scatter $\sigma_{\log M}$ (right). \textit{Bottom row}: Number of member haloes $N_{\rm members}$ (left) and ambiguity rate (right). The on-sky distribution remains approximately uniform across all selections, indicating that the quality cuts do not introduce spatial biases. The ambiguity rate histogram shows the sharp truncation at 5\% imposed by the fiducial and strict selections, while the strict selection further restricts to associations with $\sigma_R < 2.5$~Mpc and $\sigma_{\log M} < 0.25$~dex.}
    \label{fig:selection_cuts}
\end{figure}

\cref{fig:selection_cuts} illustrates the effect of these cuts on the association distributions. The on-sky distribution (top panel) remains approximately uniform across all selection levels, demonstrating that the quality cuts do not introduce spatial biases. The raw sample (grey) includes associations spanning a wide range of spatial and mass scatter, with a long tail extending to $\sigma_R > 5$~Mpc and $\sigma_{\log M} > 0.4$~dex. The fiducial selection (blue) removes associations with high ambiguity rates and low membership, retaining structures where the posterior is sufficiently sampled to characterize position and mass distributions. The strict selection (red) further restricts to the core of the scatter distributions, yielding associations with median $\sigma_R \approx 1.5$~Mpc and $\sigma_{\log M} \approx 0.15$~dex. These represent structures that are tightly localized in both position and mass within the Bayesian posterior, providing the most confident predictions from the inference.

\end{document}